\documentclass[preprint2]{aastex}

\newcommand{\Msun}{M_{\odot}}

\def\gsim{\mathrel{\rlap{\lower 4pt \hbox{\hskip 1pt $\sim$}}\raise 1pt
\hbox {$>$}}}
\def\lsim{\mathrel{\rlap{\lower 4pt \hbox{\hskip 1pt $\sim$}}\raise 1pt
\hbox {$<$}}}

\begin{document}

\title{Optical Emission from Aspherical Supernovae\\ 
and the Hypernova SN 1998bw} 

\author{
Keiichi Maeda\altaffilmark{1}, 
Paolo A. Mazzali\altaffilmark{2,3,4},
Ken'ichi Nomoto\altaffilmark{4,5} 
}

\altaffiltext{1}{Department of Earth Science and Astronomy,
Graduate School of Arts and Science, University of Tokyo, Meguro-ku, Tokyo
153-8902, Japan: maeda@esa.c.u-tokyo.ac.jp}
\altaffiltext{2}{Instituto Nazionale di Astrofisica 
(INAF)-Osservatorio Astronomico di Trieste, Via Tiepolo 11, 
I-34131 Trieste, Italy}
\altaffiltext{3}{Max-Planck-Institut f\"ur Astrophysik, 
Karl-Schwarzschild-Stra{\ss}e 1, 85741 Garching, Germany}
\altaffiltext{4}{Research Center for the Early Universe, School of
Science, University of Tokyo, Bunkyo-ku, Tokyo 113-0033, Japan}
\altaffiltext{5}{Department of Astronomy, School of Science,
University of Tokyo, Bunkyo-ku, Tokyo 113-0033, Japan}

\begin{abstract}
A fully 3D Monte Carlo scheme is applied to compute optical bolometric light 
curves for aspherical (jet-like) supernova explosion models. 
Density and 
abundance distributions are taken from hydrodynamic explosion models, 
with the energy varied as a parameter to explore the dependence. 
Our models show initially a very large degree ($\sim 4$ depending on model 
parameters) of boosting luminosity toward the polar ($z$) direction relative to 
the equatorial ($r$) plane, which decreases as the time of peak is approached. 
After the peak, the factor of the luminosity boost remains almost constant 
($\sim 1.2$) until the supernova enters the nebular phase. 
This behavior is due mostly to the aspherical $^{56}$Ni distribution in the 
earlier phase and to the disk-like inner low-velocity structure in the later 
phase. 
Also the aspherical models yield an earlier peak date than the spherical 
models, especially if viewed from near the $z$-axis. 
Aspherical models with ejecta mass $\sim 10\Msun$ are examined, 
and one with the kinetic energy of the expansion $\sim 2 \pm 0.5 \times 10^{52}$ ergs 
and a mass of $^{56}$Ni $\sim 0.4\Msun$ yields a light curve in agreement 
with the observed light curve of SN 1998bw (the prototypical hyper-energetic 
supernova).
The aspherical model is also at least qualitatively consistent with 
evolution of photospheric velocities, showing large velocities 
near the $z$-axis, and with a late-phase nebular spectrum. 
The viewing angle is close to the $z$-axis, strengthening the 
case for the association of SN~1998bw with the gamma ray burst GRB980425. 
\end{abstract}

\keywords{radiative transfer -- 
supernovae: general -- supernovae: individual (SN 1998bw) 
-- gamma rays: bursts\\
\begin{center}
\normalsize{\bf Accepted by the Astrophysical Journal.}
\end{center}
}

\section{INTRODUCTION}

Lately much emphasis has been placed on the possible role of asymmetry in 
supernova (SN) explosions, motivating a number of multi dimensional explosion
simulations both for SNe Ia (e.g., Gamezo, Khokhlov, \& Oran 2005;  R\"opke \&
Hillebrandt 2005)  and for core-collapse in massive stars  (e.g., Proga et al.
2003; Fryer \& Warren 2004;  Janka et al. 2005; Sawai, Kotake, \& Yamada 2005;
Sekiguchi \& Shibata 2005).  However, there has been surprisingly little
theoretical research on the emission from asymmetric supernovae (see e.g.,
Maeda et al. 2006a), despite the importance of this observable as a tool to test
the validity of explosion models through the comparison with observations.  In
this and subsequent papers, we report our fully 3D
computations of light curves and spectra from optical (this paper) to gamma ray
frequencies (Maeda 2006b). 

To date, only a series of papers by one group (H\"oflich 1991, 1995; 
H\"oflich, Wheeler, \& Wang 1999) has addressed two-dimensional computations of
optical light curves for aspherical supernova models. The interesting result
emerged that the optical luminosity is boosted toward the polar ($z$) direction
in a jet-like explosion because the cross section of an oblate photosphere is
larger in this direction.  They examined the early-phase light curve (before
100 days after the explosion). In the present work, we compute light curves
covering more than 1 year. 

Also, we make use of fully 3D gamma ray transport computations  for modeling
optical light curves.  To compute an optical light curve, especially at late
phases (i.e., after $\sim 100$ days), it is very important to follow correctly
the propagation of gamma rays. These are produced by radioactive decays and
penetrate the SN ejecta, depositing their energies as heat.  Previous works did
not compute optical light curves starting with 3D gamma ray transport
computations. 

For application of our light curve computations, we chose SN Ic 1998bw as a
reference. SN Ic 1998bw is the first supernova showing a direct observational
association with a gamma ray burst (GRB980425:  Galama et al. 1998), opening a
new paradigm, i.e., gamma ray burst -- supernova association (e.g., Hjorth et
al. 2003;  Kawabata et al. 2003; Matheson et al. 2003; Mazzali, et al. 2003;
Stanek et al. 2003; Deng, et al. 2005) and the origin of both as the end
product of very massive stars (MacFadyen \& Woosley 1999; Brown et al. 2000;
Wheeler et al. 2000).  Also important is that SN 1998bw itself was peculiar as 
a type Ic supernova. It was very bright, reaching a peak $V$ magnitude $\sim
-19.1$ (Galama et al. 1998, but using different distance modulus $\mu = 32.76$ 
and extinction $A_{V} = 0.1$). 
It showed very broad absorption features in the early-phase
spectra around maximum brightness.  The light curve evolution was much slower
than the well-studied SN Ic 1994I (e.g., Filippenko et al. 1995), which is
thought to be the explosion of a low mass CO star (e.g., Nomoto et al. 1994). 
These characteristics are interpreted in the context of spherically symmetric
1D models as consequences of a very energetic explosion of a massive CO star 
(Iwamoto et al. 1998; Woosley, Eastman, \& Schmidt 1999).  Iwamoto et al.
(1998) derived the kinetic energy of the explosion $E_{51} \equiv E_{\rm
K}/10^{51}$ ergs $ = 30$, the ejecta mass $M_{\rm ej} = 10.4\Msun$, the mass of
the CO star $M_{\rm CO} = 13.8\Msun$, and the main sequence mass $M_{\rm ms} =
40\Msun$. Later the value of the explosion energy was updated to be  $E_{51} =
50$ by Nakamura et al. (2001) to obtain a better fit to the optical light
curve. 

However, this was not the end of the story.  As observational data at late
epochs (e.g., Patat et al. 2001) became available, discrepancies between the
observations and the "spherical" hypernova model (e.g., Iwamoto et al. 1998)
were noticed (e.g., McKenzie \& Schaefer 1999; Sollerman et al. 2000; Mazzali
et al. 2001).  These discrepancies could not be explained simply by modifying
model parameters in the same context (e.g., Maeda et al. 2003a).  Something new
was necessary, and one promising candidate is an aspherical explosion (e.g.,
H\"oflich et al. 1999; Maeda et al. 2002).  In particular, the late-phase
spectra of SN 1998bw are well explained by an aspherical explosion 
(Maeda et al. 2002, 2006a). 

The optical light curve could also be affected by the existence of asymmetry
(H\"oflich et al. 1999; Maeda et al. 2003a).  None of the spherical
hydrodynamic models suggested to date can reproduce consistently the entire 
optical light curve of SN 1998bw, which is covered for a period of more than 1
year (see e.g., Nakamura et al. 2001).  Figure 1 shows that the synthetic light
curve for a spherical model with  $E_{51} = 50$ fits only the early phase,
while  one with $E_{51} = 10$ fits only the late phase (see \S 4 for details). 
Only {\it a priori} parameterized spherical model fits both early and late
phases  (Chugai 2000; Maeda et al. 2003a), but no studies to date have
investigated whether possible asphericity in SN 1998bw can improve the models,
in particular with regard to the late phase optical light curve. 

The aim of this paper is to provide theoretical predictions of observable
signatures of asphericity in supernovae/hypernovae, and to verify whether the
peculiarities of the prototypical hypernova SN 1998bw can be explained
consistently from early to late phases. To add consistency to our study, we not 
only examine a light curve, but also spectroscopic signatures both in early and
late phases.  The contents of the present paper are the following. 
In \S 2, we describe a direct method to compute optical and gamma ray transport 
in asymmetric supernova ejecta. Explosion models for which we examine optical 
emission are also presented.  
In \S 3, radiation transport effects in aspherical supernovae are presented, and
the differences from spherical models are discussed. 
In \S 4, synthetic light curves and their comparison to the observed light 
curve of SN 1998bw are presented. 
In \S 5, the expected photospheric velocities are compared with those of 
SN 1998bw. 
In \S 6, nebular spectra for the present models are presented. 
In \S 7, we close the paper with conclusions and discussion. 
Also in an Appendix, we describe the details of synthetic light curves, e.g., 
sensitivities on various assumptions.

\section{METHOD AND MODELS} 

\subsection{Gamma Ray and Optical Transport} 

We have developed a fully 3D, energy-dependent, and time-dependent gamma ray
transport code. We will describe details of the gamma ray transport scheme in a
subsequent paper (Maeda 2006b).  In what follows, only a brief
summary is presented. 

The code has been developed following the individual packet method 
using a Monte Carlo scheme as suggested by Lucy (2005). 
The code follows gamma ray transport in SN ejecta discretised 
in 3D Cartesian grids and in time steps. 
For the ejecta dynamics, we assume homologous expansion, which should be 
a good approximation for SNe Ia/Ib/Ic. 
The expansion of the ejecta is taken into account by dealing with the 
Doppler shift, converting at every interaction a photon packet's energy and 
direction between the comoving frame and the rest frame. 
The time delay between the emission of gamma rays and the final point (either 
absorption or escape out of the ejecta) is fully taken into account. 
Gamma ray lines from the decay chains $^{56}$Ni $\to$ $^{56}$Co $\to$ $^{56}$Fe 
and $^{57}$Ni $\to$ $^{57}$Co $\to$ $^{57}$Fe are included. 
For the interaction of gamma-rays with the SN ejecta, we consider pair 
production, Compton scattering (using the Klein-Nishina cross section), 
and photoelectric absorption (using cross sections from H to Ni). 

In view of the recent investigation by Milne et al. (2004) showing that not all 
the published 1D gamma ray transport codes give mutually consistent results, 
we tested our new code by computing gamma ray spectra based on the (spherical) 
SN Ia model W7 (Nomoto, Thielemann, \& Yokoi 1984), for which many previous 
studies are available for 
comparison. 
We compared our synthetic gamma ray spectra at 25 and 50 days with Figure 5 of 
Milne et al. (2004). We found excellent agreement between our results and the 
spectra resulting from most other codes, e.g. Hungerford, Fryer, \& Warren 
(2003). 

Starting with the detailed gamma ray transport calculations, 
we compute optical bolometric light curves. 
We again follow the scheme presented in Lucy (2005) 
(see also Cappellaro et al. 1997). 
This is a fully time-dependent computation, taking into account the time delay 
between the creation of a photon from the deposition of a gamma ray and its 
escape from the ejecta. 
The transport is solved assuming a gray atmosphere, i.e., a
frequency-independent opacity. 
We test the optical transport scheme by computing the light curve for the 
simplified SN model presented in Lucy (2005). For the test, we use a constant 
(both in space and time) opacity, same with the one used in Lucy (2005). 
We found good agreement between our light curve and Lucy (2005)'s Figure 2. 
For the optical opacity for aspherical supernova models presented in this paper, 
we phenomenologically use the formula given by Chugai (2000), 
i.e., $\kappa = 0.13/[1 + (t_{\rm d}/10)^2]$ cm$^{2}$ g$^{-1}$ where 
$t_{\rm d}$ is the time in day after the explosion.  
The formula was derived by computing opacity for the $13.8 \Msun$ CO star 
explosion model, including the contribution from electron scattering and 
bound-free transition, similar to the one used in Nakamura et al. (2001). 
By using the phenomenological formula, we miss detailed opacity distribution, 
and therefore a detailed light curve shape. Therefore, the detailed fit 
to the early phase light curve is beyond the scope of the present work 
(apparently, the late phase light curve is independent from the opacity 
prescription). This is mostly in order to save computational time (see also Lucy 2005). 

Each packet is therefore followed by the Monte Carlo method consistently 
from its creation as a gamma ray to its escape from the ejecta as either 
a gamma ray or an optical photon.  
In this study, the ejecta are mapped onto a $60^{3}$ Cartesian grid. 
In total, $5 \times 10^{6}$ photon packets (unless noted) with equal initial 
energy content are used to generate each synthetic optical light curve. 
When the packets escape from the ejecta, they are binned into 135 time steps, 
spaced logarithmically from 2 to 500 days, and into 10 angular zones with 
equal solid angle from $\theta = 0^{\rm o}$ to $180^{\rm o}$ 
(here $\theta$ is the polar angle from the $z$-axis).

\subsection{Models} 

We use Models A, C, and F of Maeda et al. (2002) to compute the optical 
emission. These are the explosions of a $16 \Msun$ He star 
(Nomoto \& Hashimoto 1988) with explosion energy 
$E_{51} \equiv E_{\rm K}/10^{51}$ ergs $ = 10$. 
At the beginning of the calculations, the energy is deposited in the sphere 
for which $M_{r} = 2.4\Msun$. The asphericity of the explosion is obtained by
depositing the energy more toward the polar direction ($z$-axis). 
Model F is a spherical model, and asphericity increases from Model C 
to Model A (see Maeda et al. 2002 for details). 

To obtain (at least qualitatively) a good fit to SN 1998bw, we examined 
several models by rescaling the ejecta density and velocity self-similarly, so 
that $M_{\rm ej} \propto \rho v^3$ and $E_{\rm K} \propto M_{\rm ej} v^2$. 
According to previous results (Iwamoto et al. 1998; Nakamura et al. 2001), we 
always used models with $M_{\rm ej} = 10.4\Msun$, which is roughly expected 
for the explosion of a $\sim 13.8\Msun$ CO star. 
 
We did not independently change the mass of $^{56}$Ni in the ejecta before  
computing light curve, since such a change would violate self-similarity 
in the synthetic light curve in terms of $E_{\rm K}$ and $M_{\rm ej}$.  
On the contrary, we obtained an approximate value of $^{56}$Ni for each 
model by scaling the flux with the original $M$($^{56}$Ni) to fit the 
observed luminosity. 
In the future we will examine light curves directly using values of 
$M_{\rm ej}$, $E_{\rm K}$, and $M$($^{56}$Ni) obtained from hydrodynamic 
simulations of explosions with various parameters. Fine-tuning 
will be necessary to obtain a nice light curve, which is time-consuming. 
For example, the initial mass cut affects the final $M$($^{56}$Ni). 
Accordingly, the initial mass cut should be fine-tuned to obtain the  
correct luminosity (see e.g., Maeda \& Nomoto 2003b). 
Figure 2 shows the ejecta structure of Model A.

\section{RADIATION TRANSPORT IN ASPHERICAL SUPERNOVAE}

Figures 3 and 4 show the last scattering points of photon packets reaching an 
observer at any direction at a given epoch for Model A with $E_{51} = 20$. 
Figure 5 shows the distribution of the total energy content of photon packets 
within position angle $\theta \sim \theta + d\theta$ as a function of the 
position angle of the points of last scattering. 
The figures were obtained from computational runs with $10^{6}$ photon packets 
(fewer than for the computations of the synthetic light curves). 
The fluctuation seen in Figure 5 is probably due to Monte Carlo noise. 

The peak date of this model is $\sim 15 - 20$ days, with some variations 
depending on the viewing angle (\S 4). 
Figures 3 and 4 show that at 10 days only photons near the surface around 
the $z$-axis escape out of the ejecta. 
Virtually no photons are emitted from the equatorial region. 
In other words, these figures consist of two sector-shaped high temperature 
emitting regions, not even of a spheroid. 
At 20 days (near the time of peak), the last scattering points move deeper as 
well as diffuse to the equatorial region, so that they are distributed at all 
angles. The innermost oblate spheroid (or butterfly) shaped region (see also 
Fig. 2) is still very optically thick and allows no photons leakage. 
At 30 days the points move even deeper. 

The peak date ($\sim 20$ days) almost coincides with the epoch when the 
distribution of last scattering points eventually covers the entire surface of
the ejecta. 
This can be understood from the distribution of the heating source $^{56}$Ni. 
Initially, the ejecta density is so high that gamma ray deposition closely 
follows the distribution of $^{56}$Ni. 
The optical photons therefore start diffusing the ejecta from the two poles. 
As time goes by, the diffusion time scale becomes smaller and smaller. 
It becomes almost equal to the expansion time scale around the time of peak. 
Therefore, photons emitted near the $z$-axis can diffuse to the equator 
around the peak date. This is unlike the usual spherically symmetric view, 
where the peak date corresponds to the date when a photon emitted near the 
center can reach the surface. 
This characteristic behavior, i.e., the photosphere moving toward the equator, 
implies that the light curve (\S 4) and the spectroscopic characteristics (\S 5)
before the peak will be differently affected by the viewing angle at different 
epochs. 

Figure 5 also shows this behavior. At epochs approaching the peak 
(i.e., between 10 and 20 days) photons diffuse to the equator. 
After that (i.e., between 20 and 30 days), the distribution of the photons' 
last scattering points does not evolve significantly. 
Note that the angular distribution of the photons is not spherically symmetric 
(as described by the sine curve overlapping with the spherically symmetric 
model F) even after the peak, but is concentrated toward the pole because of the
high temperature resulting from a larger abundance of $^{56}$Ni there.

\section{OPTICAL LIGHT CURVES}

We hereafter take the distance modulus $\mu = 32.76$ (using the Hubble constant 
$H_0 = 72$ km s$^{-1}$ Mpc$^{-1}$ and the redshift 0.0085) and the extinction 
$A_V = 0.1$ for SN 1998bw. 
Figure 1 shows optical bolometric light curves for the spherical model F.  
The model parameters are as follows: 
($M_{\rm ej}$/$\Msun$, $E_{51}$, $M$($^{56}$Ni)/$\Msun$) = 
(10.4, 10, 0.28), (10.4, 50, 0.40). 
The more energetic model ($E_{51} = 50$) yields an early peak date, 
comparable to the observed one, while the less energetic model ($E_{51} = 10$) 
peaks later than the observation. 
Although the rising part is not well fit even by the model with $E_{51} = 50$ 
in the present study, extending the $^{56}$Ni distribution out toward the 
surface would generate the rapid rise. (Note that in the present study we use 
the original $^{56}$Ni distribution obtained from the hydrodynamic computation.)
On the other hand, the late time light curve at $\sim 50 - 500$ 
days is more consistent with the less energetic model. 
After $\sim 50$ days the model with $E_{51} = 50$ declines more rapidly than 
observed. A model with $E_{51} < 10$, say $E_{51} \sim 7$ 
(Nakamura et al. 2001), would give a better fit to the late time light curve. 
This behavior of synthetic light curves is explained by simple scaling 
laws governing the peak date 
$t_{\rm peak} \propto (M_{\rm ej}^3 E_{\rm K}^{-1})^{1/4}$ 
(Arnett 1982) and the late time gamma ray optical depth 
$\tau_{\gamma} \propto M_{\rm ej}^2 E_{\rm K}^{-1}$ 
(Clocchiatti \& Wheeler 1997; Maeda et al. 2003a). 
Note that similar light curves can be obtained by varying 
the mass and energy because of the above scaling properties. 
The degeneracy can be in principal solved by 
using another information such as velocities (\S 5 and 6). 

Figure 6 shows optical bolometric light curves obtained for the aspherical 
model A. The model parameters are as follows: 
($M_{\rm ej}$/$\Msun$, $E_{51}$, $M$($^{56}$Ni)/$\Msun$) = 
(10.4, 10, 0.31), (10.4, 20, 0.39). 
The viewing angle is near the z-axis, within $37^{\rm o}$. 
Model C yields a light curve almost identical to Model A, with a slightly 
delayed peak date (1 or 2 days later than Model A for the same parameters).  

The light curve shape of the aspherical models is different from that of the 
spherical model F (Fig. 1). For given $E_{51}$, the aspherical model peaks 
earlier than the spherical model F. 
This is due to the combined effects of (1) low densities along the $z$-axis 
resulting from the large isotropic energy and (2) extended $^{56}$Ni 
distribution along the same direction. Photons therefore escape easily through 
the region near the $z$-axis (see also Figs. 3 -- 5). 
Furthermore, the aspherical models yield more photons before the peak date than 
model F, while the latter shows a rather sudden rise around the peak with no 
modification of $^{56}$Ni distribution. In a sense, the aspherical models 
naturally provide mixing of $^{56}$Ni out to the surface (toward the $z$-axis) 
as a consequence of the hydrodynamic process (Maeda et al. 2002; Maeda \& 
Nomoto 2003b), even without additional mixing processes such as Rayleigh-Taylor 
instabilities. 
At late phases, on the other hand, it resembles the spherical model 
with comparable or somewhat smaller total kinetic energy. 
For example, the aspherical model A with $E_{51} = 10$ yields a late time 
light curve flatter than the corresponding model F with $E_{51} = 10$. 
The decline rate is likely similar to the spherical model 
with $E_{51} \sim 7$ (Nakamura et al. 2001). 

As a result, the light curve is very much improved for the aspherical models as 
compared with the spherical model F (Fig. 1). 
The aspherical models reproduce better the 
observations of SN 1998bw. 
Although we did not try to optimize the fit given the uncertainty and 
simplifications in our treatment of the optical opacities (\S 2), an 
aspherical model with $E_{51} \sim 10 - 20$ gives a nice fit.

Figure 7 shows how the viewing angle affects the light curve 
for Model A with $E_{51} = 20$. 
The peak is earlier and brighter for an observer at a smaller viewing angle 
$\theta$ (angle from the $z$-axis). 
The ratio of the (apparent) luminosity for an observer on the $z$-axis to that 
for one on the equatorial plane, is close to 4 as the SN initially emerges 
around $\sim 5$ days after the explosion. The ratio then rapidly decreases 
to $\sim 1.2$ at the peak (around $\sim 20$ days).  
After the peak and up to $\sim 40$ days the ratio is almost constant and 
$\sim 1.2$, then it becomes unity as the whole ejecta become optically thin. 
For a less energetic model, the phase of constant ratio before the optically 
thin phase extends to later epochs (e.g., up to $\sim 50$ days for Model A 
with $E_{51} = 10$). 

To identify what causes this behavior we examined several models 
(see Appendix), and conclude as follows. 
Before the peak, the boost of the luminosity toward the $z$-direction is 
attributed to the fact that photons can only diffuse out to the $z$-direction 
(\S 3), i.e., that only the region around the $z$-axis is emitting photons. 
The cross sectional area of this photosphere is larger for an observer on the 
$z$-axis than on the $r$-plane, yielding a larger luminosity for the $z$-axis. 
As the peak date is approached, this region continues to expand toward the 
equator, making the luminosity difference for different directions smaller and 
smaller. After the peak, the angular distribution of the photosphere does not 
change very much any more, and the luminosity difference is almost constant. 
In this period, the central disk-like region absorbs $\gamma$-rays 
preferentially along the $z$-axis (because of the aspherical $^{56}$Ni 
distribution), and emits optical 
photons preferentially toward the $z$-direction 
(because of the disk-like structure). 

It is now interesting to compare our results to the previous work by 
H\"oflich et al (1999). 
Although the qualitative behavior, a brighter SN for smaller $\theta$, 
is mutually consistent, the details are different. 
First, our peak date depends on $\theta$ since the extended $^{56}$Ni 
distribution along the $z$-axis allows different diffusion time scales for 
different $\theta$. 
This effect was simply not included in the previous work. 
Second, before the peak the behavior is different. Our model gives initially a
very large enhancement of the luminosity toward the $z$-axis, then the effect 
gradually decreases toward the peak date. 
On the other hand, H\"oflich et al. 
(1999) gave an almost constant factor of the boost 
until the peak is reached, and after that the factor decreases 
by about a factor of 2 in 2 weeks (their Figure 6).  

The reason can be seen in the difference of the mechanism of 
boosting luminosity. 
Although in our models this is done by diffusing photons from $z$-direction to 
the equatorial direction, in H\"oflich et al. (1999) boosting luminosity is 
always attributed to an oblate spheroidal photosphere. 
Note that the input models are different. Their parameterized models 
deviate greatly from spherically symmetry, so that the axis ratio of the 
photosphere is always (from the innermost region to the outermost region) 
$\sim 2$ (see Fig. 4 of H\"oflich et al. 1999). 
In outer regions such as $v \gsim 10,000$ km s$^{-1}$ (depending on $E_{51}$) 
the density does not largely deviate from spherical symmetry in our models. 
Only below this the density contour shows a disk like structure (Fig. 2), 
but with a smaller axial ratio than theirs. 
Finally, in our model the difference in luminosities for different $\theta$ is 
much smaller than theirs after the peak. 
In this phase, the boost of the luminosity is attributed to both the aspherical 
$^{56}$Ni distribution and the disk-like density structure (see above), while 
in H\"oflich et al. (1999) it is due only to the disk-like structure. 
The degree of the boost is smaller for our model than H\"oflich et al. (1999) 
partly because the deviation of the ejecta density distribution from 
spherical symmetry is smaller in our models, although the direct comparison is 
difficult because the mechanism is different. 
Also, the low angular resolution for small $\theta$ in the present calculations 
(i.e., averaging the interval $0^{\rm o} \sim 37^{\rm o}$) may also smooth the 
luminosity difference to some extent. This is due to the current choice of an 
equal solid angle for photon binning, which is most efficient in the 
Monte-Carlo scheme. In the future, we will use finer bins with a larger number 
of photon packets. In any case, this resolution effect is probably small. 

Of course, which models are more realistic is a different question.  Although
our models are based on hydrodynamic computations (in this sense our models are
at least more self-consistent), different parameters for explosion 
calculations may lead to density structure more similar to H\"oflich et al. 
(1999). 
Actually, most of the differences may be understood in terms of 
the different progenitors (the ejecta mass $\sim 2$ and $10\Msun$, 
for H\"oflich et al. (1999) and the present study, respectively). 
One may expect larger asphericity in the inner region for low mass cores. 

In any case, we emphasize that it is very important to treat the 
$^{56}$Ni distribution correctly and the gamma ray transport in 3D, as this is 
in the present model the main cause of the characteristic behavior of the light
curves. Also, we would suspect based on our result and analysis that the naive
guess that larger asphericity (by a jet powered explosion) always yields a
larger luminosity boost toward the $z$-axis may not be correct. For example, if
$^{56}$Ni-rich blobs penetrate entirely through the star as an extreme case of
a jet powered explosion, then the SN could be brighter near the $r$-plane than
on the $z$-axis because two blobs are seen from the $r$-plane but just one from
the $z$-axis.  In our model the blobs are stopped in the star and therefore the
SN is brighter on the $z$-axis.  In any case, light curve behavior is highly
sensitive to density and $^{56}$Ni distribution, and therefore one should
always be careful when deriving asphericity (if possible) from the light curve.
One should always check the consistency with other information, e.g.,
spectroscopic characteristics, as we proceed to do in subsequent sections.

\section{PHOTOSPHERIC VELOCITIES}

A model can be further constrained from its spectroscopic characteristics. 
For example, the evolution of the photospheric velocity provides 
important information since it scales roughly
as $v_{\rm ph} \propto \sqrt{E_{\rm K}/M_{\rm ej}}$, while 
the light curve shape depends on these parameters differently (\S 4). 

The concept of the photosphere needs careful reconsideration if the ejecta are
not spherically symmetric.   Deriving the photospheric velocity from observed
spectra, in fact, implicitly assumes that the ejecta are spherically
symmetric.  An absorption minimum corresponds to the line-of-sight velocity of 
a slice (perpendicular to an observer) where the amount of absorption takes a
maximum value. In the spherically symmetric case, this velocity is exactly what
is called the photospheric velocity, but in an asymmetric geometry this may not
be the case.  Therefore, direct comparison between a multi dimensional model
and observations ultimately needs multi dimensional (early phase) spectrum
synthesis computations (e.g., Kasen et al. 2004), which are beyond the scope of
the present paper. 

We give a very rough estimate of the photospheric velocity by looking at the
line-of-sight velocity of the last scattering points of optical photons emitted
toward a given observer's direction.  Figure 8 shows the photons' energy
content distribution as functions of the line-of-sight velocity of the last
scattering points and of the epoch.  The brightest part gives a rough estimate
of the photospheric velocity.  By definition the photospheric velocity in
spherically symmetric ejecta is the line-of-sight velocity of the (nearly) last
scattering point moving toward the observer, while Figure 8 shows the
line-of-sight velocity of "all" the last scattering points from which photons
are emitted in the  direction to the observer. Therefore, taking into account
photons emitted aside the line connecting the observer and the center of the SN
ejecta, this distribution will probably somewhat underestimate the 
photospheric velocity. Also, the velocity is apparently sensitively affected by
the optical opacity, therefore uncertainty in the opacity leads to  
uncertainty in the photospheric velocity. 

Given the above caveat,  Figure 8 shows that the spherical model F with $E_{51}
= 50$ gives a photospheric velocity consistent with SN 1998bw. Model F with the
smaller energy ($E_{51} = 10$) gives a velocity smaller than observed.  
The aspherical model A yields different "photospheric" velocities for different
directions as well as different luminosities for different directions. 
Especially before the peak there is a large velocity boost toward
the $z$-axis compared to the corresponding spherical model with the same
energy. Indeed, Model A with $E_{51} = 20$ gives a velocity for the $z$-axis
before the peak similar to that of Model F with $E_{51} = 50$. 

Figure 8 shows that Model A with $E_{51} = 20$ is consistent with SN 1998bw. 
The model A with $E_{51} = 10$ gives a bit smaller velocity before $\sim 15$
days.  However, it should be stated that in Model A with $E_{51} = 10$, the
maximum velocity in the ejecta is $\sim 20,000$ km s$^{-1}$. Because the
decreasing density near the surface is difficult to follow correctly in 2D/3D
hydrodynamic simulations, we might be underestimating the density at $\gsim
20,000$ km s$^{-1}$.  After the peak the photosphere moves deeper and this
uncertainty is not a problem.  If we judge the model photospheric velocity at
the peak, even Model A with $E_{51} = 10$  may marginally be consistent with
the observation.  Therefore, the aspherical model A with $E_{51} = 10 \sim 20$
is consistent with the observed photospheric velocities, assuming the 
observer is placed near the $z$-direction.

\section{NEBULAR SPECTRA}

After the ejecta become optically thin,  the SN evolves into the nebular
phase.  The emission process is now different from the earlier photospheric
phase. For the nebular phase a 3D spectrum synthesis method has already been
developed (Kozma et al. 2005; Maeda et al. 2006a).  We compute the nebular
emission at 350 days after the explosion for the present models.  The detail of
the nebular emission computation in 3D and its application to SN 1998bw are
presented in Maeda et al. (2006a). 
In Maeda et al. (2006a), the
mass of some elements are modified independently to obtain a detailed fit,
while in this work we do not allow this. 
Because in this paper we rescale the
hydrodynamic models somewhat differently, 
we again compute nebular spectra for the present models to test consistency. 

Figure 9 shows synthetic nebular spectra for the present models at $350$ days
after the explosion as compared with the observed spectrum of SN 1998bw at
$337$ days after the peak  (Patat et al. 2001).  Although no attempt has been
made to obtain a good fit to SN 1998bw, the aspherical model A is at least
qualitatively consistent with the observation.  For example, no spherical
models give the sharply peaked [OI] 6300, 6363\AA\ doublet, the strongest line,
as is observed (e.g., Sollerman et al. 2000; Mazzali et al. 2001).  On the
other hand, the aspherical model A with $E_{51} = 10 \sim 20$ yields the
sharply peaked [OI] line, with a viewing angle close to the $z$-axis as is
consistent with the light curve (\S 4), with photospheric
velocity analysis (\S 5), and with the detection of a GRB.  See Maeda et al.
(2002, 2006a) for details. 
The observed nebular line broadness agrees well with Model A with 
$E_{51} = 10$, but is a bit narrower than the synthetic lines of 
Model A with $E_{51} = 20$. 
This may possibly imply a less massive progenitor for SN 1998bw 
than examined in the present study ($M_{\rm ms} = 40\Msun$), which is 
further discussed in \S 7.

\section{CONCLUSIONS AND DISCUSSION}

The aims of this paper are (1) to examine the optical properties of aspherical
supernovae, including light curves covering more than $1$ year, early phase
photospheric velocities, and late phase nebular spectra, and (2) to investigate
if one can obtain a model consistent with hypernova SN 1998bw in all these
aspects.  We have presented observational signatures expected from aspherical
hypernova models of Maeda et al. (2002) and closely compare them with
observations of SN 1998bw. 

We use a Monte Carlo scheme to solve the time-dependent radiation transport
problem in homologously expanding SN ejecta in 3D grids.  It is demonstrated
that the scheme is very suitable to the problem.  For example, the concept of
the last scattering points, which can easily be traced in the Monte Carlo
scheme, turns out to be very useful to understand the physical reasons of some
non-trivial behaviors in time-dependent 3D radiation transport (see e.g.,
Appendix). 

The light curves of the aspherical models show characteristics different from 
those of spherical models. The aspherical model A (or C) of Maeda et al. (2002) 
shows both (1) early emergence of optical emission  and (2) boosted luminosity
toward the $z$-axis before the peak. These are mainly attributed to
concentration of $^{56}$Ni along the $z$-axis.  The degree of boosting on the
$z$-axis relative to the $r$-axis decreases with time, from a factor of $\sim
4$ to $\sim 1.2$.  After the peak and until the nebular phase, the boosting is
smaller than at earlier phases. In this phase, the ratio of the luminosity in
the $z$-axis to that on the $r$-plane remains almost constant at $\sim 1.2$.  
The photospheric velocity is dependent on the viewing angle,  and is higher on
the $z$-axis than on the $r$-plane.  Model A with $E_{51} = 20$ gives 
photospheric velocities comparable to those of the spherical model F with
$E_{51} = 50$.  Nebular spectra are also different for different asphericity
and different viewing angle.  If the aspherical model is viewed on the
$z$-axis, it yields sharply peaked O and Mg lines, and broad Fe lines (see
Maeda et al. 2006a for more detail). 

All these features are consistent with the observations of SN 1998bw. Note 
that these properties could not be explained in spherically symmetric models. 
The highly aspherical model (A or C in Maeda et al. (2002)) with $M_{\rm ej}
\sim 10\Msun$,  $M$($^{56}$Ni) $\sim 0.4\Msun$, and $E_{51} = 10 \sim 20$ 
gives a nice fit to  all these observations. Strictly speaking, there is still a
small difficulty, i.e.,  the early phase observations favor the larger energy
$E_{51} = 20$ (\S 4 and 5)  although the late phase observations favor the
smaller energy $E_{51} = 10$ (\S 4 and 6).  However, given the fact that we use
the "hydrodynamic" model, the situation is  much better than in the spherical
models (i.e., $E_{51} = 7$ vs.  $E_{51} = 50$; see Nakamura et al. 2001).  For
example, if Model A with $E_{51} = 10$ has more high velocity materials with
some $^{56}$Ni  at $V > 20,000$ km s$^{-1}$ than found in the hydrodynamic
simulation (see discussion in \S 5), then the problems in the early phase (a
somewhat late peak and smaller photospheric velocity than observed) may be
resolved.  This will lead to an increased expansion kinetic energy, and $E_{51}
\sim 20 \pm 5$ may be a solution. 
Another possibility is that the progenitor may be less massive than examined in 
the present study. A less massive star tends to be compact in inner regions 
and to be extended in outer regions, so that it may yield 
average velocities higher in the early phases (i.e., outer parts) 
and lower in the late phases (i.e., inner parts) than a more massive star. 
However, it is unlikely that the progenitor's mass is very different from 
$M_{\rm ms} = 40\Msun$ since then the photospheric velocity will not follow 
the observations. In any case, it is an interesting possibility and should be 
addresses in future studies. Apparently, to do this it is necessary to 
compute hydrodynamics of the explosions for various progenitors rather than 
scaling the mass and energy, since the different density distribution will be 
an important factor.

The mass of $^{56}$Ni $\sim 0.4\Msun$ is consistent with the result 
of the aspherical explosion model of Maeda et al. (2002), where initial 
mass cut is set at $M_{r} = 2.4\Msun$ and all the $^{56}$Ni is produced by 
explosive nucleosynthesis in the shock waves propagating the progenitor star. 
If the initial mass cut (initial remnant mass) is larger, 
another mechanism such as a massive disk wind (MacFadyen 2003) is necessary. 
In either case, the nucleosynthesis products are suggested to be different 
from conventional spherically symmetric models (e.g., Nagataki 2000; 
Maeda et al 2002; Maeda \& Nomoto 2003b; Prute, Surman, \& McLaughlin 2004a; 
Pruet, Thompson, \& Hoffman 2004b). 
Some elements such as Zn, Ti are suggested to be enhanced. This could have a 
very important consequence on the Galactic chemical evolution (e.g., Kobayashi 
et al. 2006), especially at earliest phases (e.g., Iwamoto et al. 2005). 
These nucleosynthetic features could be examined 
by observing abundance patterns of either extremely metal poor halo stars 
(e.g., Christlieb et al. 2002; Frebel et al. 2005) 
or binary systems having experienced a supernova explosion of the primary star 
(e.g., Podsiadlowski et al. 2002; Gonz\'alez Hern\'andez et al. 2005). 

A viewing angle $\theta < 30^{\rm o}$ is found to be  consistent with, indeed
even favorable for, the observed optical light curve of SN 1998bw.  
This is consistent with the result of H\"oflich et al. (1999).
The result seems
very natural given the detection of a gamma ray burst in association with SN
1998bw.  This is also consistent with a somewhat "off-axis" jet model ($\theta
\sim 15^{\rm o}$;  Yamazaki, Yonetoku, \& Nakamura 2003) suggested for SN
1998bw. Placing tighter constraints on the viewing angle from models of the
optical emission is unfortunately difficult, since the optical emission comes
from  non-relativistic ejecta which do not show relativistic boosting. 

Comparison of our models with other supernovae, including hypernovae SN 1997ef
and  2002ap, should be interesting.  Late time light curves of SNe 1997ef
(e.g., Mazzali, Iwamoto, \& Nomoto 2000) and 2002ap (e.g., Mazzali et al.
2002)  are also inconsistent with spherical models (see e.g., Maeda et al.
2003a). To solve the problem in these SNe will need extensive survey of $M_{\rm
ej}$, $E_{51}$, $M$($^{56}$Ni), and the ejecta geometry (the degree of
asphericity). Qualitatively, they deviate from the prediction of spherically
symmetric model in the same manner  as SN 1998bw did, so that asphericity is a
promising candidate to solve the discrepancy.  If it is, then asphericity is a
general property of hypernovae.  
Indeed, the asphericity is likely a general feature of core-collapse 
supernovae. 
There are some supernovae,
probably of normal energy, that possibly show a similar light curve behavior 
(e.g., Clocchiatti et al. 1996; Clocchiatti \& Wheeler 1997).  
A direct {\it Hubble Space Telescope} image of SN 1987A clearly 
shows bipolarity (Wang et al. 2002). 
Also, polarization and spectropolarization measurements suggest that core-collapse 
supernovae are essentially asymmetric (Wang et al. 2001; Leonard, Filippenko, \& Ardila 2001).
If available,
modeling not only light curves, but also nebular spectra (e.g., Matheson et al.
2001)  is very important to constraint the nature of the explosion  (see also
Mazzali et al. 2005) as we see in the present paper. 

Also interesting is the high energy emission from aspherical
supernovae/hypernovae.  It should be noted that computations of high energy
emission  based on the present model give model predictions "consistent" with
optical observations of hypernova SN 1998bw.  This consistent modeling between
optical and high energy emissions  has only been done for the very nearby SN
1987A (e.g., McCray, Shull, \& Sutherland 1987; Woosley et al. 1987; Shibazaki,
\& Ebisuzaki 1988; Kumagai et al. 1989).  Because a hypernova is potentially a
very interesting event not only in optical but also in $X$  and $\gamma$ rays
because of a large amount of radioactivity (especially $^{56}$Ni:  see e.g.,
Nomoto et al. 2004) and large kinetic energy, model predictions consistent with
the optical observations should be provided. We will present $X$ and $\gamma$
ray emission for the present models in a subsequent paper (Maeda 2006b).

\acknowledgements

K.M is supported through the JSPS 
(Japan Society for the Promotion of Science) 
Research Fellowship for Young Scientists. 
The authors would like to thank 
Jinsong Deng, Nozomu Tominaga, and Masaomi Tanaka 
for useful discussion, and Ferdinando Patat for 
the observational data of SN 1998bw.  

\appendix

\section{DETAILS OF SYNTHETIC LIGHT CURVES}

In this section, we examine several models based on Model A with $E_{51} = 20$. 
We examine how the following factors affect the synthetic light curves. 
(1) Distribution of $^{56}$Ni. 
(2) Treatment of time duration of a photon packet spent in the ejecta. 

Taking Model A with $E_{51} = 20$, 
(1) we assume that all the $^{56}$Ni is at the center, while the density structure is unchanged 
(we call this case "central $^{56}$Ni"). 
This is an approximation to the case where the $^{56}$Ni is always deeper than 
the photosphere. 
(2) We assume that the diffusion time has no angular dependence. 
We further assume $L_{\theta} (t) = f(\theta, t) L_{\rm mean} (t)$, 
where the mean luminosity $L_{\rm mean} (t)$ is assumed to be that of 
the corresponding "spherical" model with the same 
$M_{\rm ej}$, $M$($^{56}$Ni), and $E_{51}$, and 
$f (\theta, t)$ is the angular distribution of emitted photons at 
fixed time $t$ (assuming that the photon diffusion time at any angle 
is equal to that of the spherical model).  
These assumptions in the case (2) are used in the previous work by H\"oflich et al. (1999) 
(we call this case "Approximate diffusion"). 

Following three cases are examined. 
(a) The central $^{56}$Ni distribution, using the "correct" angle dependent diffusion time 
fully computed with a time-dependent 3D computation ("central $^{56}$Ni"), 
(b) the original $^{56}$Ni distribution, using the non-angle dependent diffusion time 
as described above ("Approximate diffusion"), and 
(c) the central $^{56}$Ni and non-angle dependent diffusion, which is probably 
most similar to the case  
examined in H\"oflich et al. (1999) ("$^{56}$Ni centered + Approximate diffusion"). 
Figures 10 and 11 show the angular luminosity distribution and light curves, respectively. 
Figures 12 -- 14 show the cross sectional view (on the $V_{x} - V_{z}$ plane) of the last 
scattering points for optical photon packets. 

How the $^{56}$Ni distribution affects the light curve appearance is seen by 
comparing Figures 7a and 7b (with original aspherical distribution) with 
Figures 10a and 11a (with $^{56}$Ni centered distribution). 
The effects of treatment of diffusion time is seen by comparing 
Figures 7a and 7b [10a and 11a] (directly solving 3D transport) with Figures 10b and 11b 
[10c and 11c] (with the approximate diffusion). 

First, we found that the aspherical distribution of $^{56}$Ni, extending toward
the surface  along the $z$-axis, is important to make the early emergence of
optical photons before the peak. All three models show the evolution of the
light curves around the peak  much slower than the original model (\S 4). Note
that even Case b using  the original $^{56}$Ni distribution for computing the
angle dependence is in effect similar to the $^{56}$Ni centered models (a and
c), because of the assumption $L_{\rm mean} = L_{\rm spherical}$ where the
spherical model shows more or less centered $^{56}$Ni distribution. Also
interesting is that the models with the original $^{56}$Ni distribution
(original and Case b) show the boost of luminosity toward the $z$-axis before
the peak, while the ones with the centered $^{56}$Ni distribution  (Cases a and
c) do not. Note that while Case a apparently shows a luminosity boost  before
the peak in Figure 10a, this effect disappears well before the peak  so that
the effect is not seen in Figure 11a. A close view of the last scattering
points (Figures 4 and 12 -- 14) suggests that before the peak the boost of the
luminosity  toward the $z$-axis can mainly be attributed to the aspherical
$^{56}$Ni distribution and energy deposition (with minor contribution due to
low densities and small diffusion time scale along the $z$-axis). 

Second, we found that the treatment of the diffusion time scale is also very
important.  Especially we point out that the assumption that $L_{\rm mean} (t)$
is equal to the corresponding "spherical" model is not always correct.  It
ignores the fact that the $^{56}$Ni distribution is different for the
aspherical and spherical models. For example, comparing the original model with
Case b demonstrates that the "mean" luminosity computed for the original model
is not similar to the corresponding spherical model. 

Finally, all the three cases as well as the original model (\S 4) show  an
almost constant factor of boosting luminosity after the peak, and this effect 
is especially evident for Case a and c ($^{56}$Ni centered).  A close view of
the last scattering points illustrates that for Cases a and c at these epochs
the photosphere is an oblate or disk-like spheroid with
larger cross sectional area for smaller viewing angle $\theta$ as suggested by
H\"olich et al (1999).  However, the original model and Case b with the
original aspherical $^{56}$Ni  distribution show the distribution of the last
scattering points preferentially along the $z$-axis.  If the $^{56}$Ni is
distributed along the $z$-axis, $\gamma$-rays  are absorbed preferentially
along the $z$-axis, making temperature  higher toward the $z$-axis. Because
this high temperature spots are on the disk-like structure, they emit more
toward the $z$-axis than the $r$-axis. 

From these test calculations, we conclude that the behavior of the synthetic
light curve of  our aspherical models can be understood as follows.  First, the
angular dependence can be divided into two phases, i.e., before and after the
peak.  Before the peak, boosting luminosity toward the $z$-axis is mainly
attributed to the  aspherical $^{56}$Ni distribution.  After the peak, it is
caused by combined effects of  the aspherical $^{56}$Ni distribution and the
disk-like inner structure.  The fact that all the three models give light
curves different from the original model A, especially in reproducing the rapid
rise before the peak, demonstrates the importance  of direct time dependent
computations of light curves without crude approximation on  the transport
processes.

\onecolumn

\clearpage
\begin{figure}
\begin{center}
	\begin{minipage}[t]{0.4\textwidth}
		\epsscale{1.0}
		\plotone{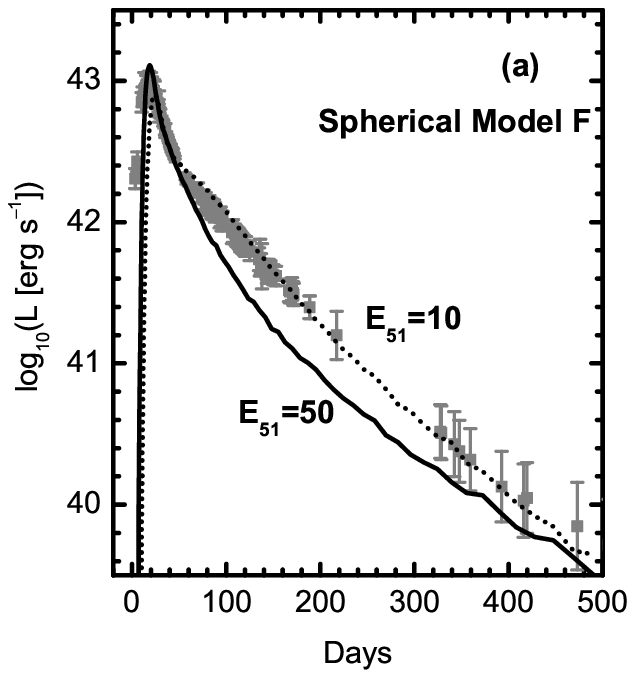}
	\end{minipage}
	\begin{minipage}[t]{0.4\textwidth}
		\epsscale{1.0}
		\plotone{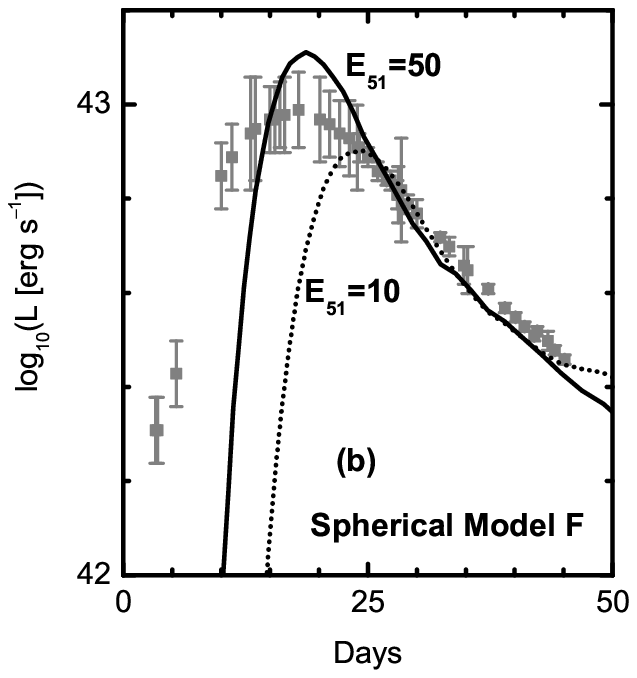}
	\end{minipage}
\end{center}
\caption[]
{Light Curves for the spherical model F 
(Maeda et al. 2002) with the kinetic energy 
of the expansion $E_{51} \equiv E_{\rm K}/10^{51} = 50$ (solid) 
and $10$ (dotted). 
The synthetic curves are obtained with $10^{6}$ photon packets. 
The bolometric UVOIR light curve of SN 1998bw is taken from 
Patat et al. (2001), with the distance modulus $\mu = 32.76$ 
and the extinction $A_V = 0.1$. 
The left panel (a) shows the light curve up to $500$ days, 
while the right panel (b) shows that up to only $50$ days. 
\label{fig1}}
\end{figure}

\clearpage
\begin{figure}
\begin{center}
	\begin{minipage}[t]{0.4\textwidth}
		\epsscale{1.0}
		\plotone{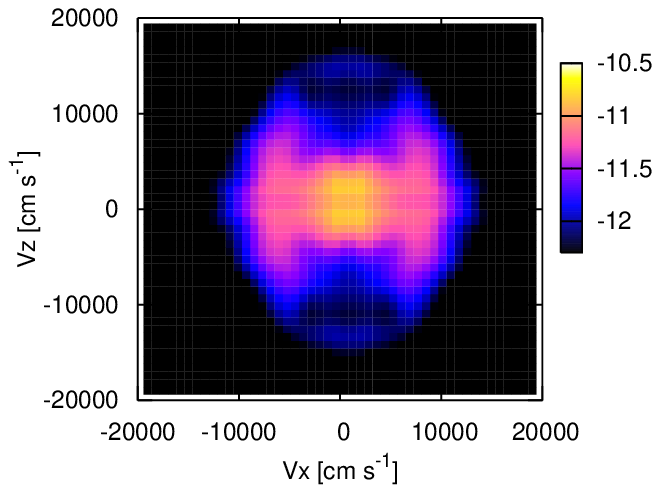}
	\end{minipage}
	\begin{minipage}[t]{0.4\textwidth}
		\epsscale{1.0}
		\plotone{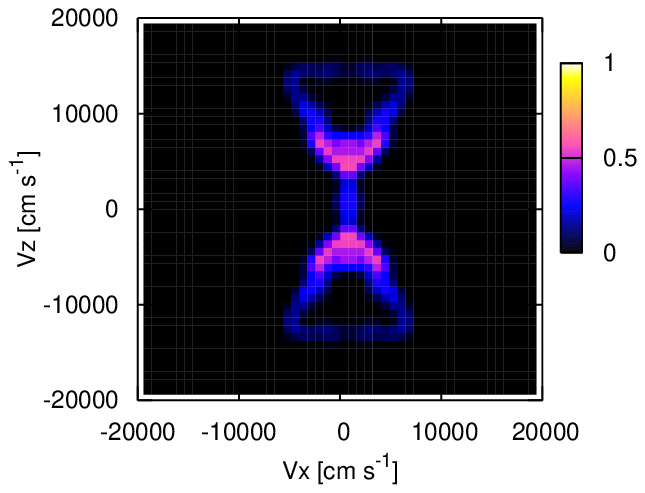}
	\end{minipage}
\end{center}
\caption[]
{Model A (Maeda et al. 2002) with $E_{51} = 10$. 
The left panel shows density distribution in a logarithmic scale 
($\log(\rho ~[{\rm g~cm}^{-3}])$) at 10 days after the explosion. The ejecta is already in a 
homologous expansion phase, so that the distribution is shown in 
the velocity space. 
The right panel shows 
mass fractions of $^{56}$Ni in a linear scale. 
\label{fig2}}
\end{figure}

\clearpage
\begin{figure}
\begin{center}
	\hspace{-10cm}
	\begin{minipage}[t]{0.4\textwidth}
		\epsscale{2.5}
		\plotone{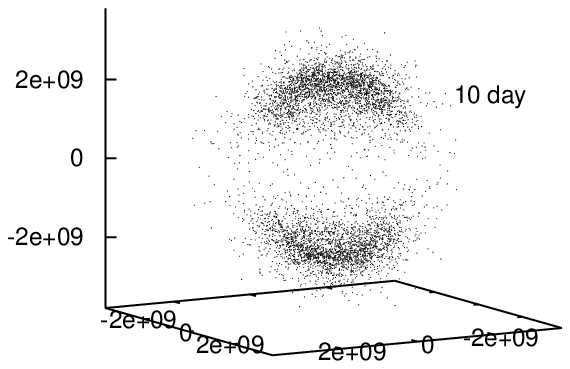}
	\end{minipage}
	\begin{minipage}[t]{0.4\textwidth}
		\epsscale{2.5}
		\plotone{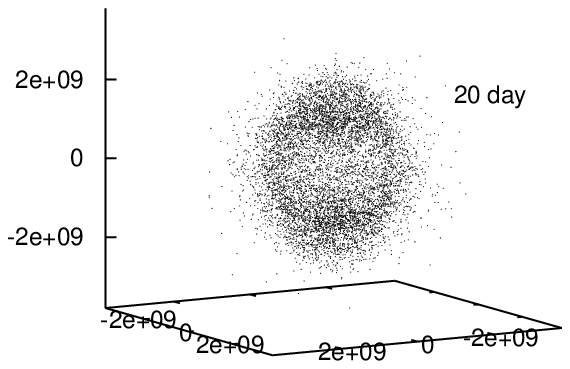}
	\end{minipage}\\
\end{center}
\caption[]
{Last scattering points of optical photon packets 
escaping out of the ejecta in Model A with $E_{51} = 20$ 
at $10$ (left) and $20$ (right) days after the explosion. 
The points are shown in the velocity space ($V_{x}$,$V_{y}$,$V_{z}$ 
[cm s$^{-1}$]) and the vertical axis is the $V_{z}$ axis. 
The total number of optical photons packets is $10^{6}$. 
\label{fig3}}
\end{figure}

\clearpage
\begin{figure}
\begin{center}
	\begin{minipage}[t]{0.4\textwidth}
		\epsscale{1.0}
		\plotone{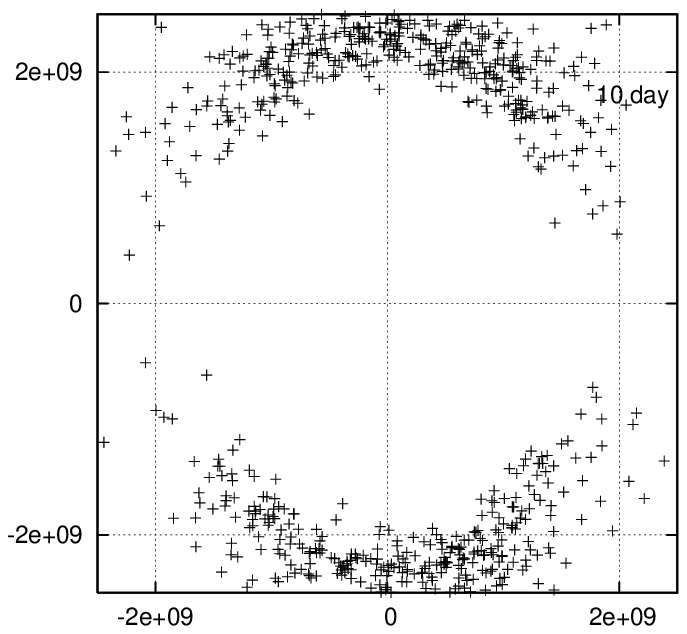}
	\end{minipage}
	\begin{minipage}[t]{0.4\textwidth}
		\epsscale{1.0}
		\plotone{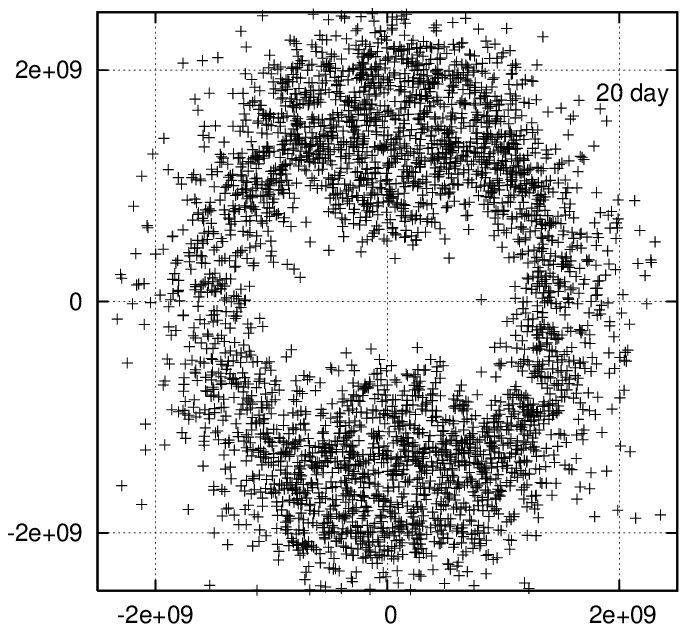}
	\end{minipage}\\
	\begin{minipage}[t]{0.4\textwidth}
		\epsscale{1.0}
		\plotone{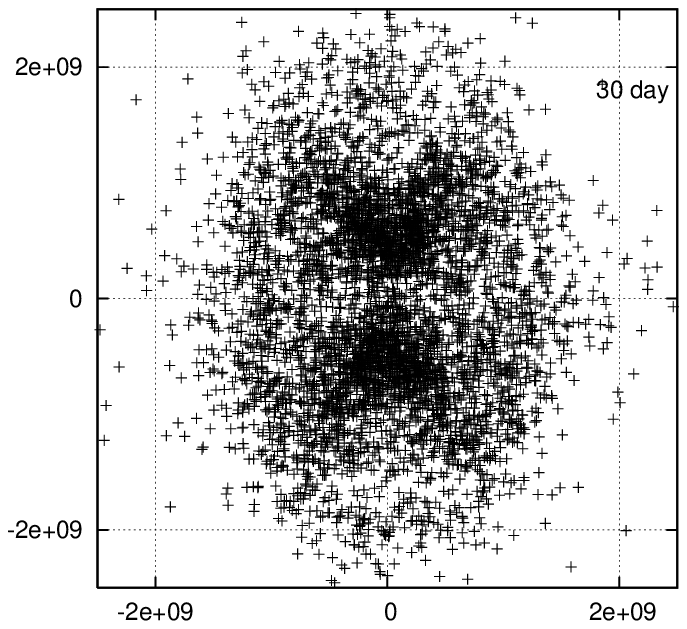}
	\end{minipage}
	\begin{minipage}[t]{0.4\textwidth}
		\epsscale{1.0}
		\plotone{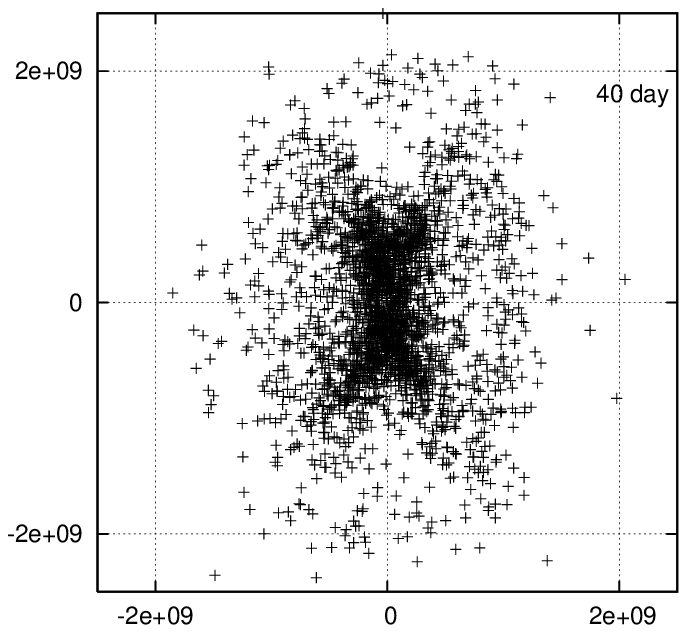}
	\end{minipage}
\end{center}
\caption[]
{Last scattering points of optical photon packets 
escaping out of the ejecta in Model A with $E_{51} = 20$ 
shown in the $V_{x} - V_{z}$ plane (i.e., the slice at 
$V_{y} = 0$). The vertical and horizontal axes are 
respectively the $V_{z}$ and the $V_{x}$ axes. 
The last scattering points with $-2 
< V_{y}/10^{8} {\rm cm \ s}^{-1} < 2$ are shown. The distribution is shown 
for $10$, $20$, $30$, and $40$ days after the explosion. 
\label{fig4}}
\end{figure}

\clearpage
\begin{figure}
\begin{center}
	\begin{minipage}[t]{1.0\textwidth}
		\epsscale{0.5}
		\plotone{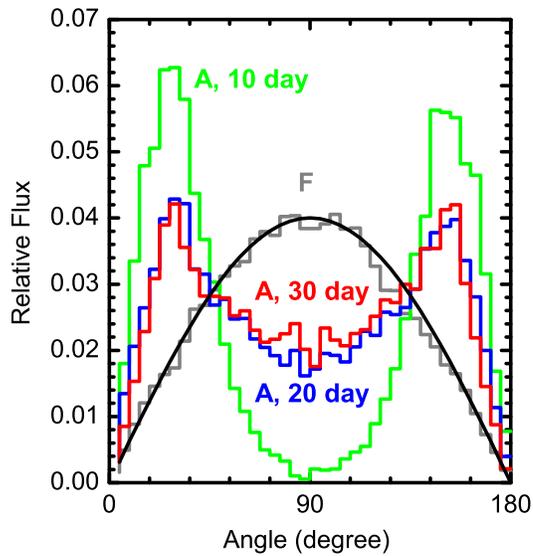}
	\end{minipage}
\end{center}
\caption[]
{Angular distribution of the optical photons' total energy content. 
The energy content of the optical photons (arbitrary scaled) 
is binned according to the polar angle $\theta$ of the last 
scattering point within the equal angle $4.5^{\rm o}$ each. 
Therefore, any spherically symmetric models should ideally give 
a sine curve (black). 
The distribution is shown for Model A at 10 (green), 
20 (blue), and 30 days (red) 
after the explosion and for Model F (gray). 
The number of photons used in the computation is $10^{6}$. 
\label{fig5}}
\end{figure}

\clearpage
\begin{figure}
\begin{center}
	\begin{minipage}[t]{0.4\textwidth}
		\epsscale{1.0}
		\plotone{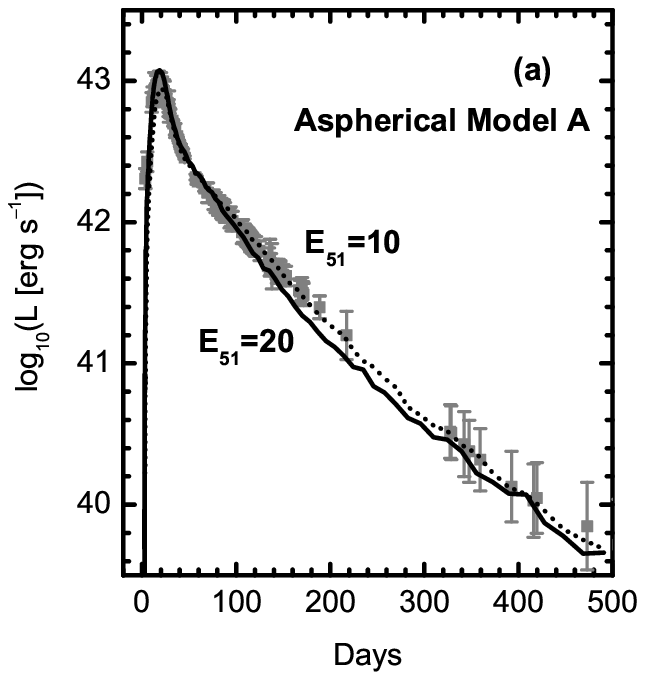}
	\end{minipage}
	\begin{minipage}[t]{0.4\textwidth}
		\epsscale{1.0}
		\plotone{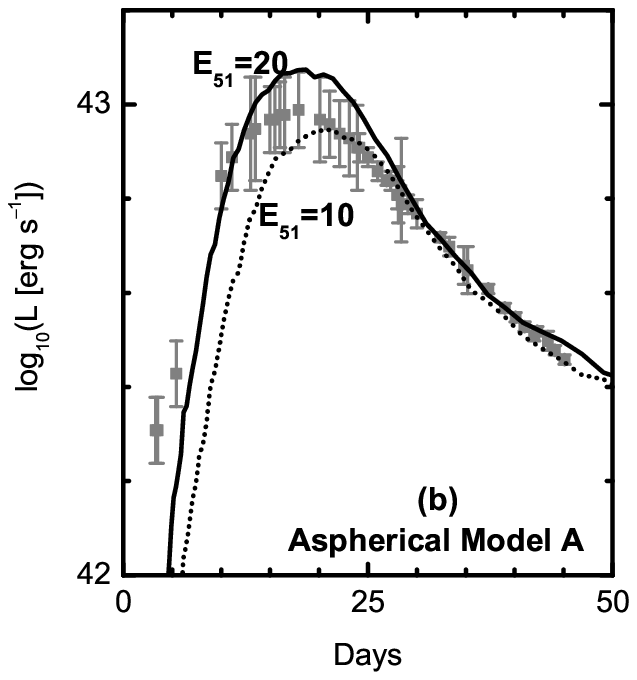}
	\end{minipage}
\end{center}
\caption[]
{Light Curves for the aspherical model A 
(Maeda et al. 2002) with the kinetic energy 
of the expansion $E_{51} \equiv E_{\rm K}/10^{51} = 20$ (solid) 
and $10$ (dotted). The observer's direction is along the $z$-axis 
within $\theta < 37^{\rm o}$. 
The synthetic curves are obtained with $5 \times 10^{6}$ photon packets. 
The left panel (a) shows the light curve up to $500$ days, 
while the right panel (b) shows that up to only $50$ days. 
\label{fig6}}
\end{figure}

\clearpage
\begin{figure}
\begin{center}
	\begin{minipage}[t]{0.4\textwidth}
		\epsscale{1.0}
		\plotone{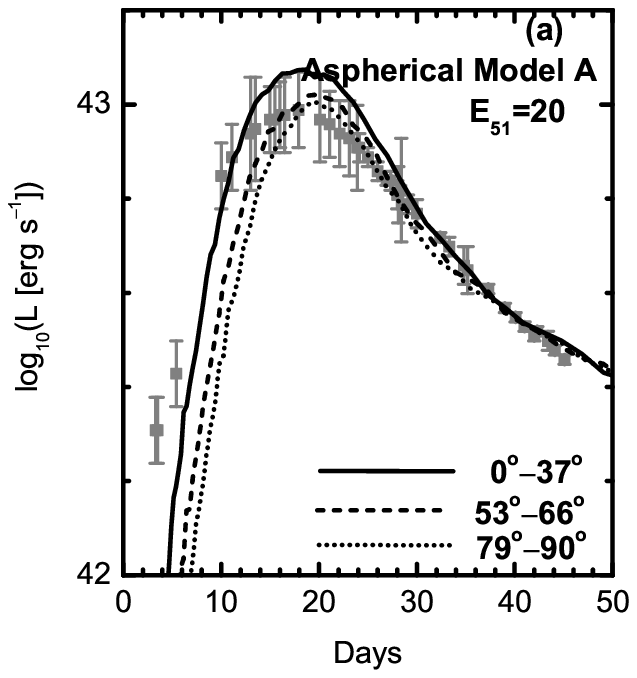}
	\end{minipage}
	\begin{minipage}[t]{0.4\textwidth}
		\epsscale{1.0}
		\plotone{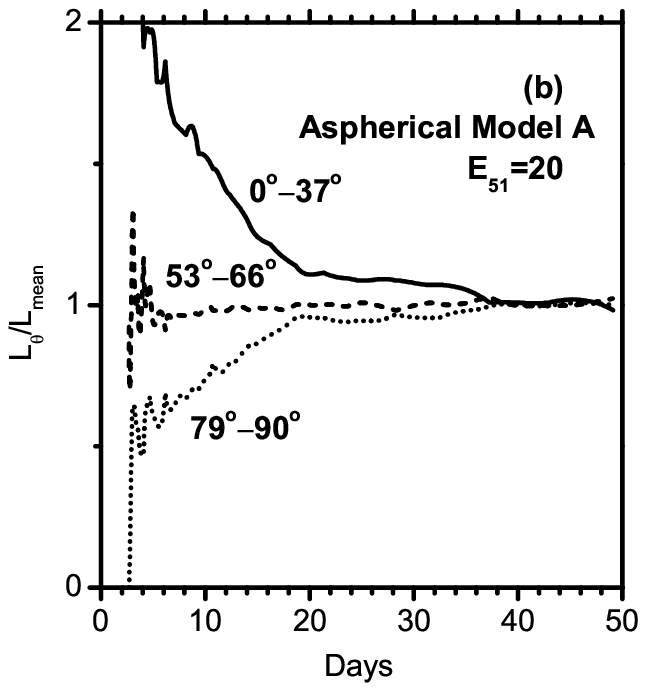}
	\end{minipage}
\end{center}
\caption[]
{Dependence of the synthetic light curve on the viewing angle $\theta$ 
in Model A with $E_{51} = 20$. 
The left panel (a) shows the synthetic light curves for 
different observer's directions (solid for $0^{\rm o} < \theta < 37^{\rm o}$, 
dashed for $53^{\rm o} < \theta < 66^{\rm o}$, and 
dotted for $79^{\rm o} < \theta < 90^{\rm o}$). 
The synthetic curves are obtained with $5 \times 10^{6}$ photon packets. 
The right panel (b) shows the luminosities at different $\theta$ normalized 
by the mean luminosity $L_{\rm mean}$. Here $L_{\rm mean}$ is the luminosity 
averaged over all the solid angles (from $0^{\rm o}$ to $180^{\rm o}$).  
\label{fig7}}
\end{figure}

\clearpage
\begin{figure}
\begin{center}
	\begin{minipage}[t]{0.4\textwidth}
		\epsscale{1.75}
		\plotone{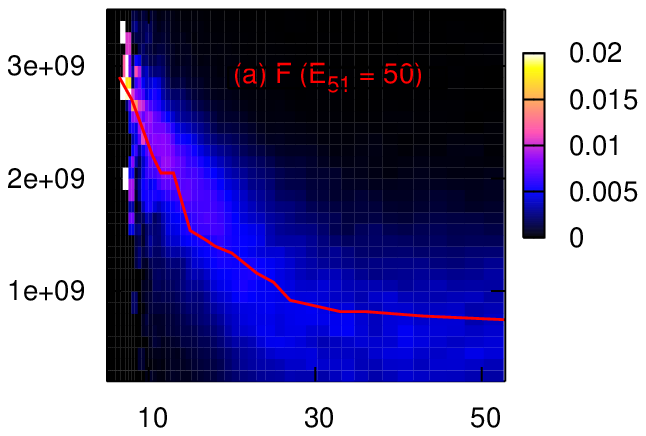}
	\end{minipage}
	\begin{minipage}[t]{0.4\textwidth}
		\epsscale{1.75}
		\plotone{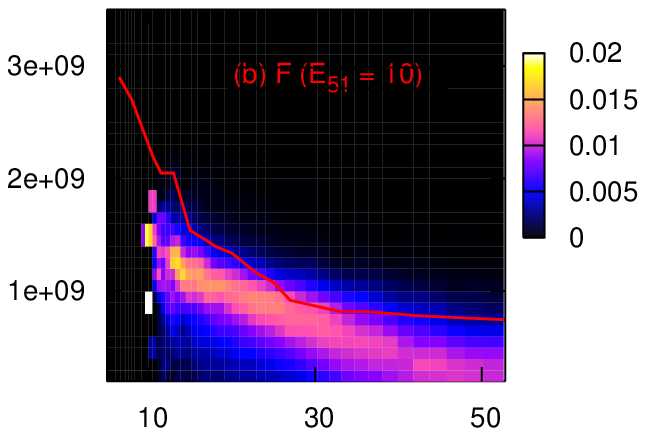}
	\end{minipage}\\
	\vspace{-4cm}
	\begin{minipage}[t]{0.4\textwidth}
		\epsscale{1.75}
		\plotone{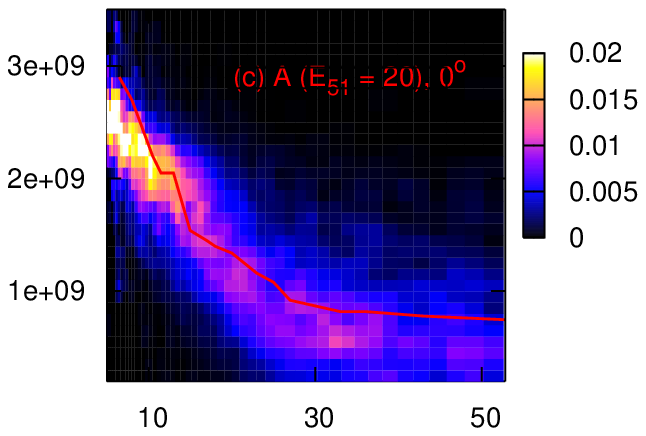}
	\end{minipage}
	\begin{minipage}[t]{0.4\textwidth}
		\epsscale{1.75}
		\plotone{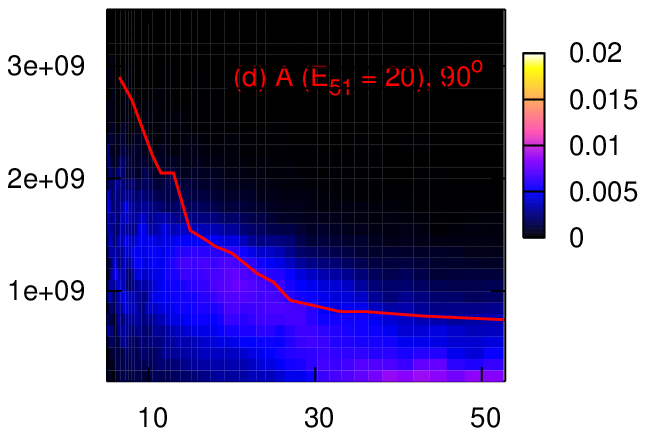}
	\end{minipage}\\
	\vspace{-4cm}
	\begin{minipage}[t]{0.4\textwidth}
		\epsscale{1.75}
		\plotone{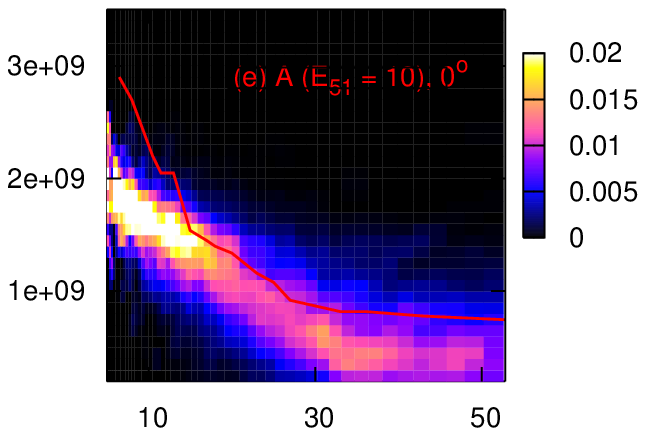}
	\end{minipage}
	\begin{minipage}[t]{0.4\textwidth}
		\epsscale{1.75}
		\plotone{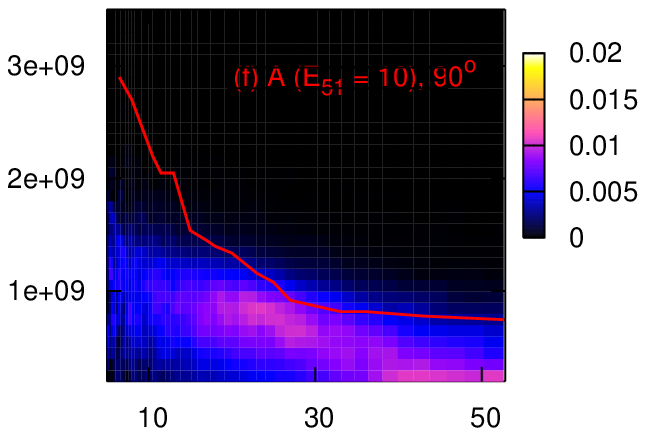}
	\end{minipage}\\
\end{center}
\caption[]
{Time evolution (horizontal axes) of 
distribution of photon's energy contents as a function of 
line-of-sight velocities (vertical axes) at the last scattering points. 
The bright regions give a rough estimate of photospheric velocities 
as a function of time. See \S 5 for details. The distribution is shown for 
(a) the spherical model F with $E_{51} = 50$, (b) F with $E_{51} = 10$, 
(c) the aspherical model A with $E_{51} = 20$ and $\theta = 0^{\rm o}$, 
(d) A with $E_{51} = 20$ and $\theta = 90^{\rm o}$, 
(e) A with $E_{51} = 10$ and $\theta = 0^{\rm o}$, and 
(c) A with $E_{51} = 10$ and $\theta = 90^{\rm o}$. 
Here $\theta$ is the polar angle (from the $z$-axis) of the observer's direction.  
Also shown is the photospheric velocity of SN 1998bw as a function of time 
(curve; Patat 2001). 
\label{fig8}}
\end{figure}

\clearpage
\begin{figure}
\begin{center}
	\begin{minipage}[t]{0.4\textwidth}
		\epsscale{1.0}
		\plotone{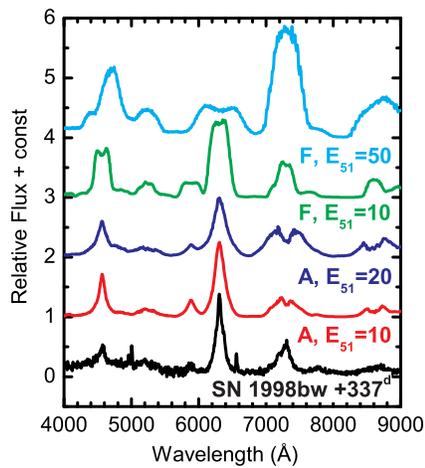}
	\end{minipage}
\end{center}
\caption[]
{Synthetic nebular spectra (at 350 days after the explosion) 
for Model F with $E_{51} = 50$ (cyan), F with $E_{51} = 10$ (green), 
Model A with $E_{51} = 20$ (blue), and A with $E_{51} = 10$ (red). 
The viewing angle is $30^{\rm o}$ from the $z$-axis for Model A.  
Also shown is the observed spectrum of SN 1998bw at 337 days after the maximum 
brightness (black). 
\label{fig9}}
\end{figure}

\clearpage
\begin{figure}
\begin{center}
	\begin{minipage}[t]{0.4\textwidth}
		\epsscale{1.0}
		\plotone{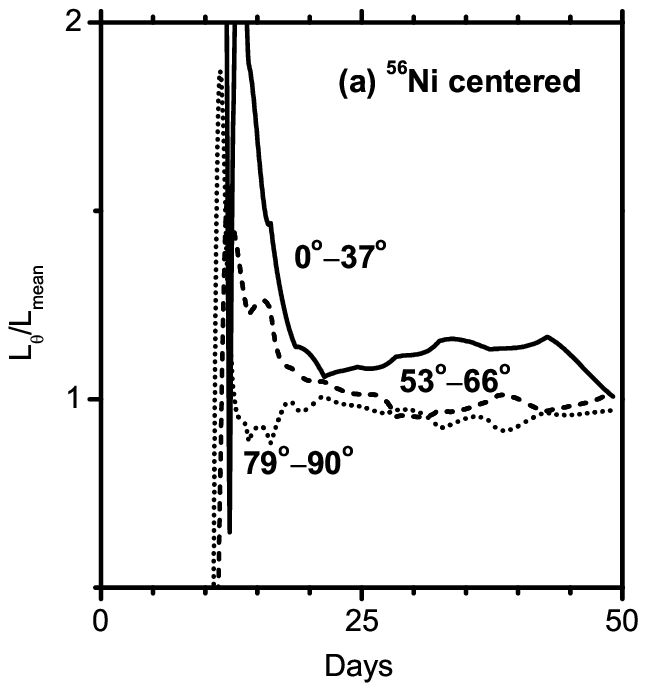}
	\end{minipage}
	\begin{minipage}[t]{0.4\textwidth}
		\epsscale{1.0}
		\plotone{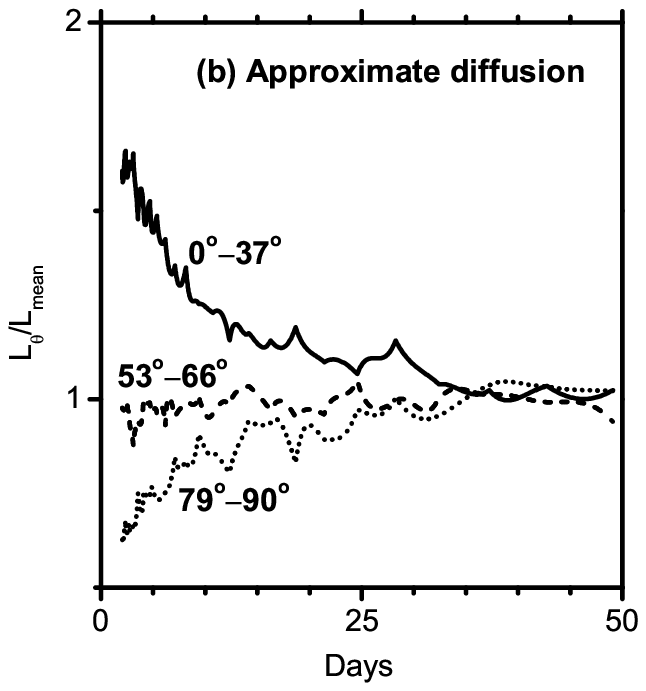}
	\end{minipage}
	\begin{minipage}[t]{0.4\textwidth}
		\epsscale{1.0}
		\plotone{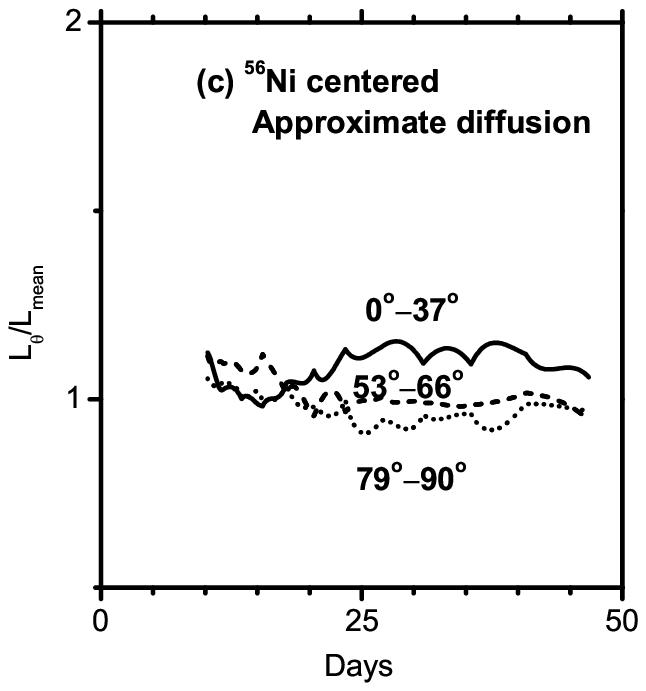}
	\end{minipage}
\end{center}
\caption[]
{Synthetic light curves normalized by the mean luminosity $L_{\rm mean}$, 
depending on the viewing angle $\theta$ 
(solid for $0^{\rm o} < \theta < 37^{\rm o}$, 
dashed for $53^{\rm o} < \theta < 66^{\rm o}$, and 
dotted for $79^{\rm o} < \theta < 90^{\rm o}$) 
for Model A with $E_{51} = 20$. 
Here $L_{\rm mean}$ is the luminosity 
averaged over all the solid angles (from $0^{\rm o}$ to $180^{\rm o}$).  
Three cases are shown: 
The panel (a) shows the model with $^{56}$Ni distribution 
(artificially) concentrated at the center ("$^{56}$Ni centered"). 
The panel (b) shows the model with the assumptions  
$L_{\rm mean} = L_{\rm spherical}$ (a spherical model with the same 
energy) and no-angle dependent diffusion time 
("Approximate diffusion"; See Appendix for details). 
The panel (c) shows the model with all the assumptions in (a) and (b). 
The models are computed with $10^{6}$ photon packets. 
\label{fig10}}
\end{figure}

\clearpage
\begin{figure}
\begin{center}
	\begin{minipage}[t]{0.4\textwidth}
		\epsscale{1.0}
		\plotone{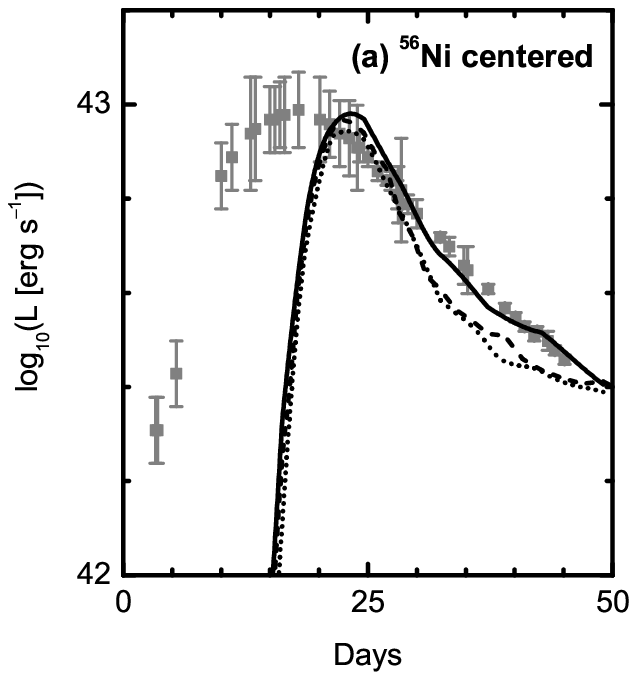}
	\end{minipage}
	\begin{minipage}[t]{0.4\textwidth}
		\epsscale{1.0}
		\plotone{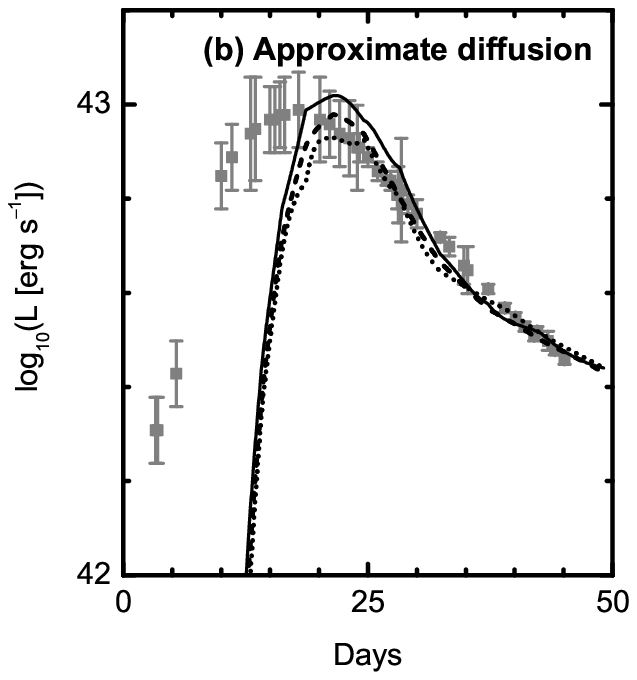}
	\end{minipage}
	\begin{minipage}[t]{0.4\textwidth}
		\epsscale{1.0}
		\plotone{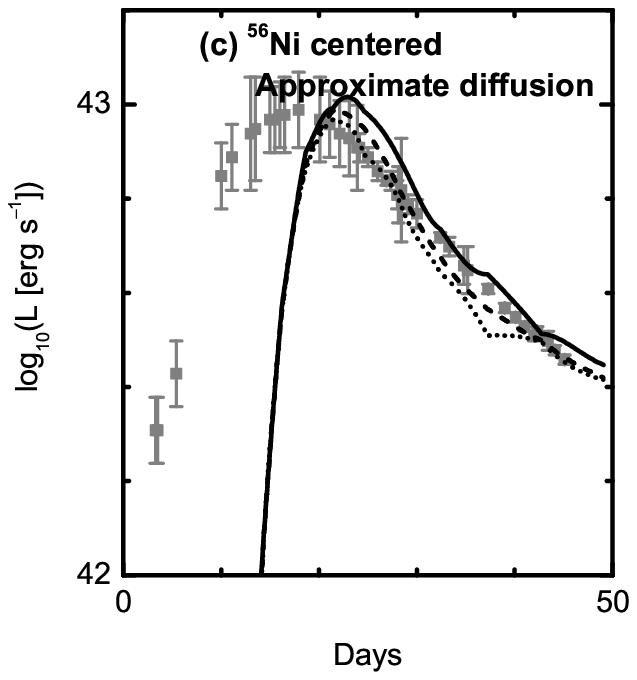}
	\end{minipage}
\end{center}
\caption[]
{Synthetic light curves for the model A with $E_{51} = 20$ 
with various assumptions, i.e., (a) "$^{56}$Ni centered", 
(b) "Approximate diffusion", and (c) "$^{56}$Ni centered and  
Approximate diffusion". See the caption of Figure 10 for details.
The masses of $^{56}$Ni in the models 
are $0.27\Msun$ (Case a) and $0.29\Msun$ 
(Cases b and c) (Note, however, that in Appendix we do not 
attempt to derive the mass of $^{56}$Ni). 
 \label{fig11}}
\end{figure}

\clearpage
\begin{figure}
\begin{center}
	\begin{minipage}[t]{0.4\textwidth}
		\epsscale{1.0}
		\plotone{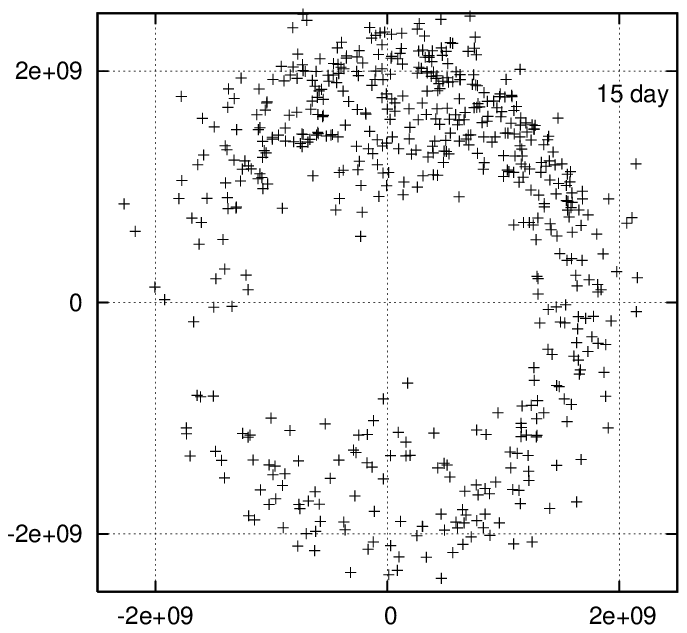}
	\end{minipage}
	\begin{minipage}[t]{0.4\textwidth}
		\epsscale{1.0}
		\plotone{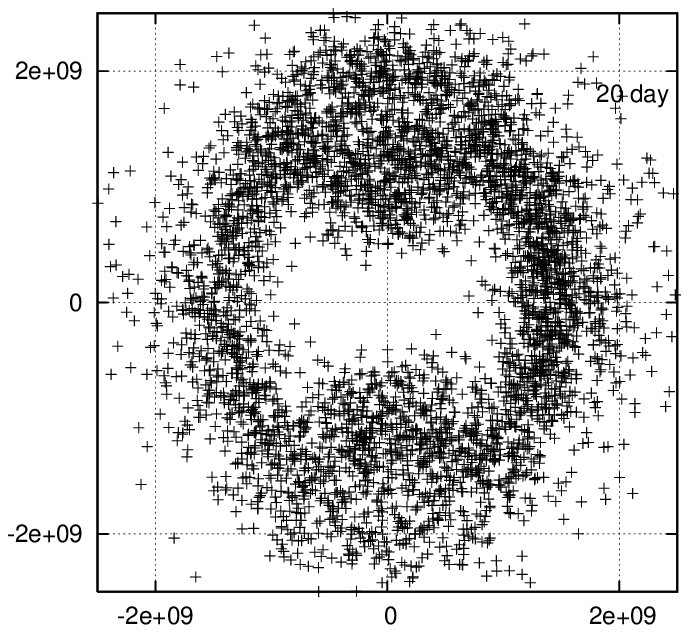}
	\end{minipage}\\
	\begin{minipage}[t]{0.4\textwidth}
		\epsscale{1.0}
		\plotone{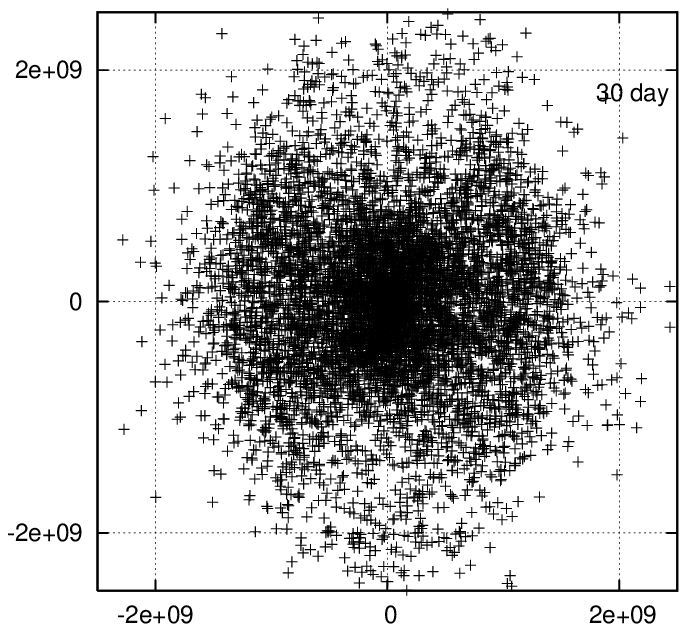}
	\end{minipage}
	\begin{minipage}[t]{0.4\textwidth}
		\epsscale{1.0}
		\plotone{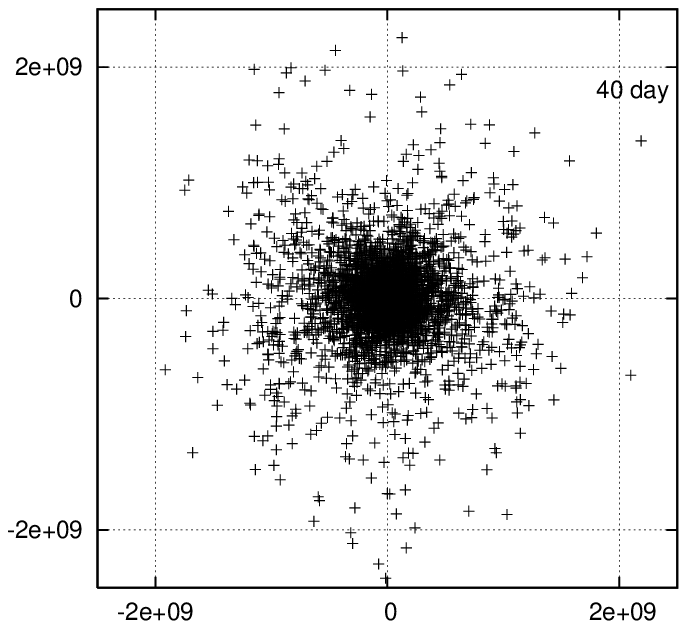}
	\end{minipage}
\end{center}
\caption[]
{Last scattering points for Model A with $E_{51} = 20$, with 
$^{56}$Ni distribution (artificially) concentrated at the 
center ("$^{56}$Ni centered"). 
Last scattering points of optical photon packets  
are shown in the $V_{x} - V_{z}$ plane (i.e., the slice at 
$V_{y} = 0$). The vertical and horizontal axes are 
respectively the $V_{z}$ and the $V_{x}$ axes. 
The last scattering points with $-2 
< V_{y}/10^{8} {\rm cm \ s}^{-1} < 2$ are shown. The distribution is shown 
for $15$, $20$, $30$, and $40$ days after the explosion. 
\label{fig12}}
\end{figure}

\clearpage
\begin{figure}
\begin{center}
	\begin{minipage}[t]{0.4\textwidth}
		\epsscale{1.0}
 		\plotone{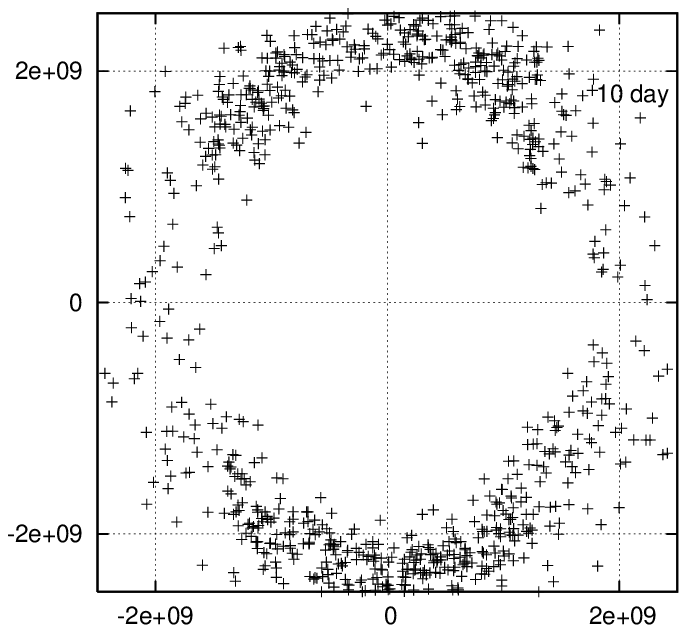}
	\end{minipage}
	\begin{minipage}[t]{0.4\textwidth}
		\epsscale{1.0}
		\plotone{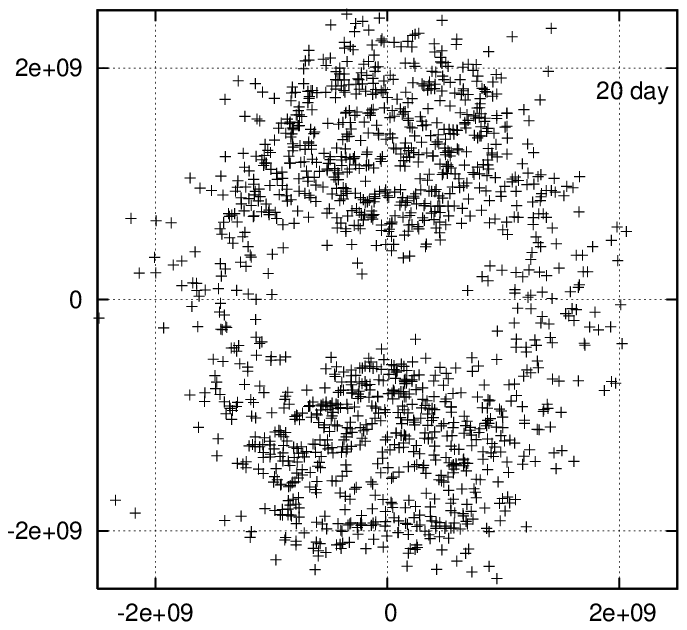}
	\end{minipage}\\
	\begin{minipage}[t]{0.4\textwidth}
		\epsscale{1.0}
		\plotone{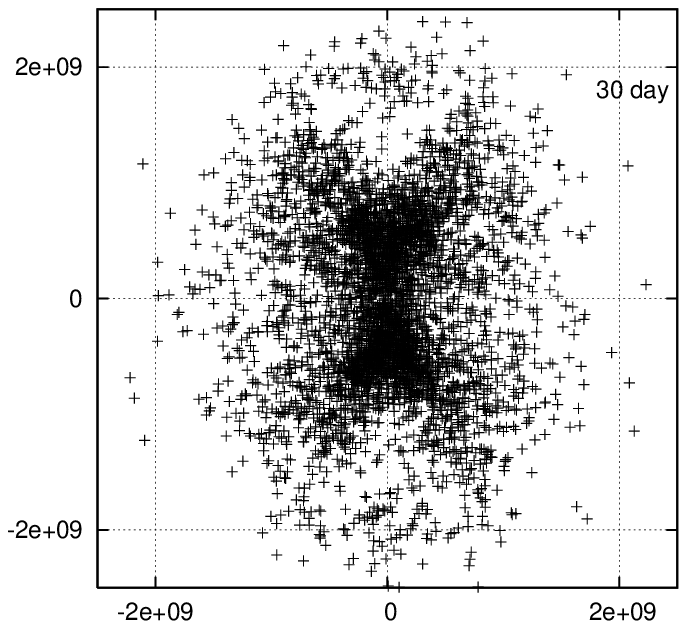}
	\end{minipage}
	\begin{minipage}[t]{0.4\textwidth}
		\epsscale{1.0}
		\plotone{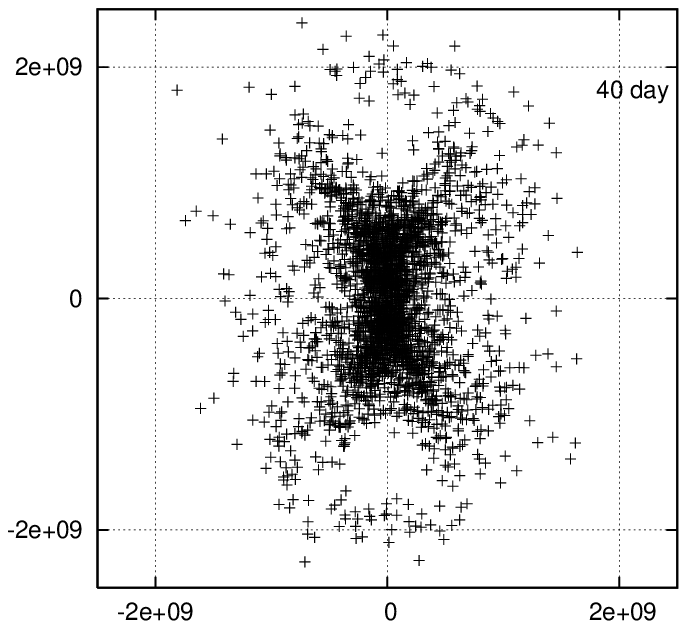}
	\end{minipage}
\end{center}
\caption[]
{Last scattering points for Model A with $E_{51} = 20$, with 
the assumptions  
$L_{\rm mean} = L_{\rm spherical}$ and no-angle dependent diffusion time 
("Approximate diffusion"; See Appendix for details). 
Last scattering points of optical photon packets  
are shown in the $V_{x} - V_{z}$ plane (i.e., the slice at 
$V_{y} = 0$). The vertical and horizontal axes are 
respectively the $V_{z}$ and the $V_{x}$ axes. 
The last scattering points with $-2 
< V_{y}/10^{8} {\rm cm \ s}^{-1} < 2$ are shown. The distribution is shown 
for $10$, $20$, $30$, and $40$ days after the explosion. 
\label{fig13}}
\end{figure}

\clearpage
\begin{figure}
\begin{center}
	\begin{minipage}[t]{0.4\textwidth}
		\epsscale{1.0}
 		\plotone{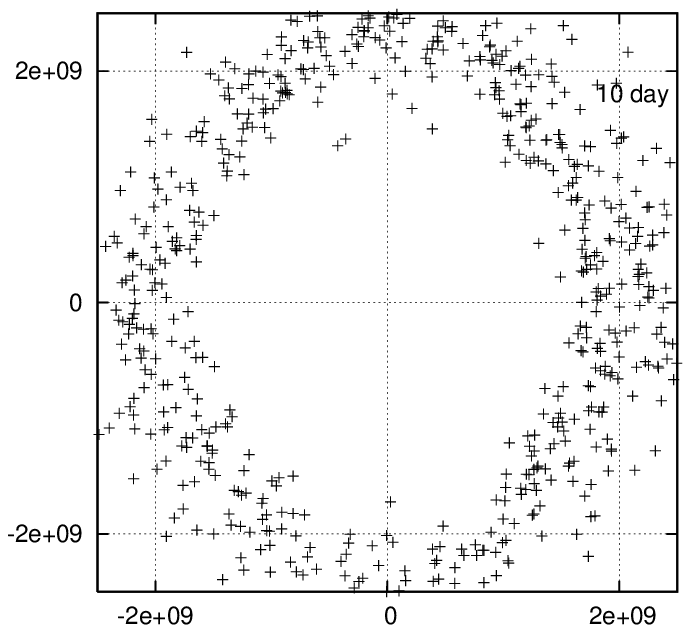}
	\end{minipage}
	\begin{minipage}[t]{0.4\textwidth}
		\epsscale{1.0}
		\plotone{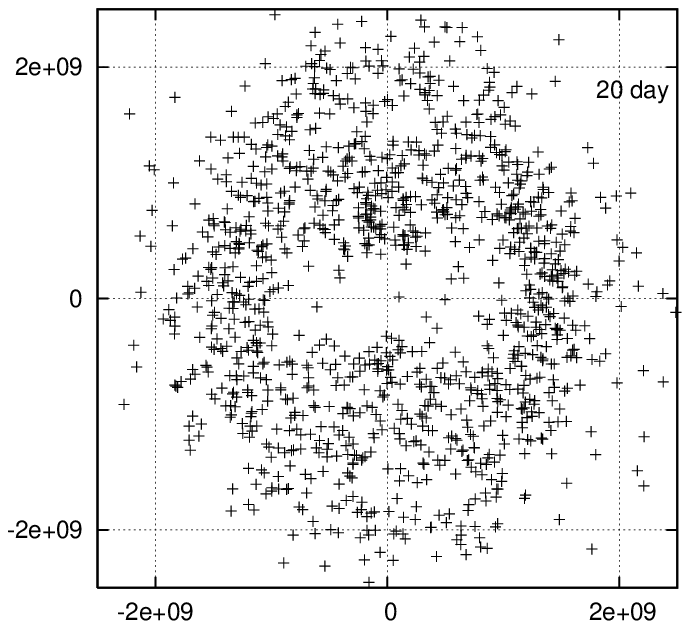}
	\end{minipage}\\
	\begin{minipage}[t]{0.4\textwidth}
		\epsscale{1.0}
		\plotone{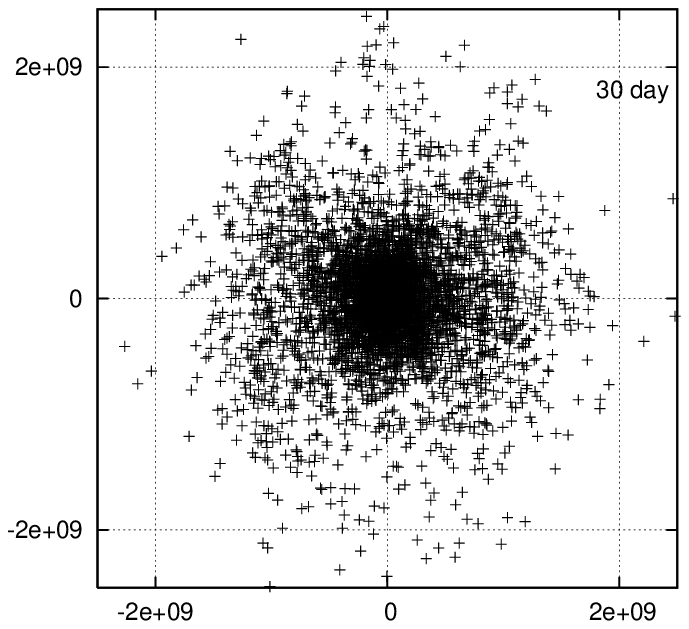}
	\end{minipage}
	\begin{minipage}[t]{0.4\textwidth}
		\epsscale{1.0}
		\plotone{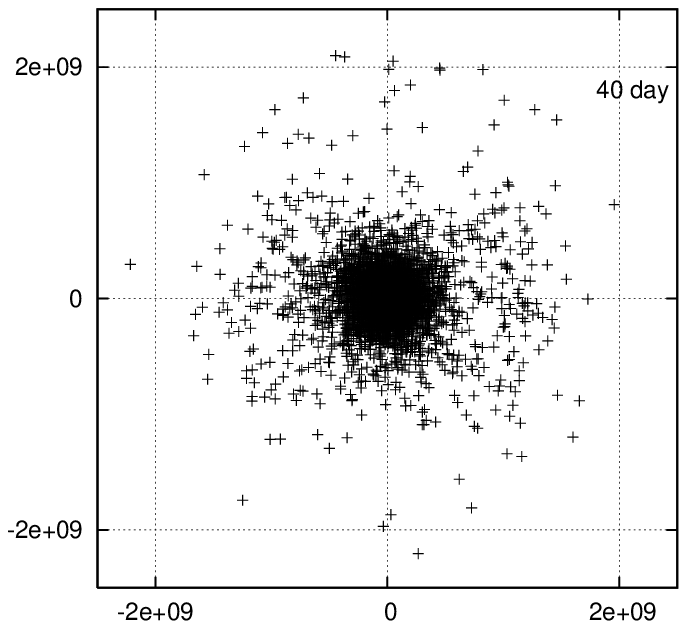}
	\end{minipage}
\end{center}
\caption[]
{Last scattering points for Model A with $E_{51} = 20$ with 
$^{56}$Ni distribution (artificially) concentrated at the center, 
with the assumptions  
$L_{\rm mean} = L_{\rm spherical}$ and no-angle dependent diffusion time 
("$^{56}$Ni centered" and "Approximate diffusion"; See Appendix for details). 
Last scattering points of optical photon packets  
are shown in the $V_{x} - V_{z}$ plane (i.e., the slice at 
$V_{y} = 0$). The vertical and horizontal axes are 
respectively the $V_{z}$ and the $V_{x}$ axes. 
The last scattering points with $-2 
< V_{y}/10^{8} {\rm cm \ s}^{-1} < 2$ are shown. The distribution is shown 
for $10$, $20$, $30$, and $40$ days after the explosion. 
\label{fig14}}
\end{figure}


\begin{thebibliography}{99}

\bibitem[]{809}
Arnett, W.D. 1982, ApJ, 253, 785

\bibitem[]{812}
Brown, G.E., Lee, C.-H., Wijers, R.A.M.J., Lee, H.K., Israelian, G., 
\& Bethe, H.A. 
2000, New Astronomy, 5, 191

\bibitem[]{817}
Cappellaro, E., Mazzali, P.A., Benetti, S., Danziger, I.J., 
Turatto, M., Della Valle, M., \& Patat, F. 
1997, A\&A, 328, 203

\bibitem[]{822}
Christlieb, N., et al. 2002, Nature, 419, 904 

\bibitem[]{825}
Clocchiatti, A., Wheeler, J.C., Benetti, S., \& Frueh, M. 1996, 
ApJ, 459, 547

\bibitem[]{829}
Clocchiatti, A., \& Wheeler, J.C. 1997, ApJ, 491, 375

\bibitem[]{832}
Chugai, N.N. 2000, Astronomy Letters, 26, 797

\bibitem[]{835}
Deng, J., Tominaga, N., Mazzali, P.A., Maeda, K., 
\& Nomoto, K. 2005, ApJ, 624, 898

\bibitem[]{839}
Filippenko, A.V., et al. 1995, ApJ, 450, L11

\bibitem[]{842}
Frebel, A., et al. 2005, Nature, 434, 871 

\bibitem[]{845}
Fryer, C.L., \& Warren, M.S. 2004, ApJ, 601, 391

\bibitem[]{848}
Galama, T.J., et al. 1998, Nature, 395, 670

\bibitem[]{851}
Gamezo, V. N., Khokhlov, A.M., 
\& Oran, E.S. 2005, ApJ, 623, 337

\bibitem[]{855}
Gonz\'alez Hern\'andez, J.I., Rebolo, R., 
Israelian, G., Casares, J., Maeda, K., 
Bonifacio, P., \& Molara, P. 
2005, ApJ, 630, 495

\bibitem[]{861}
Hjorth, J., et al. 2003, Nature, 423, 847

\bibitem[]{864}
H\"oflich, P.  1991, A\&A, 246, 481

\bibitem[]{867}
H\"oflich, P. 1995, ApJ, 440, 821

\bibitem[]{870}
H\"oflich, P., Wheeler, J.C., \& Wang, L. 1999, ApJ, 521, 179

\bibitem[]{873}
Hungerford, A.L., Fryer, C.L., 
\& Warren, M.S. 2003, ApJ, 594, 390

\bibitem[]{877}
Iwamoto, K., et al. 1998, Nature, 395, 672

\bibitem[]{880}
Iwamoto, N., Umeda, H., Tominaga, N., Nomoto, K., 
\& Maeda, K. 2005, Science, 309, 451

\bibitem[]{884}
Janka, H.-Th., Scheck, L., Kifonidis, K., M\"uller, E., \& Plewa, T. 
2005, in "The Fate of the Most Massive Stars", ASP Conference Proceedings, 
(eds. P. Humphreys and K. Stanek), vol. 332, 372

\bibitem[]{889}
Kasen, D., Nugent, P., Thomas, R.C., \& Wang, L. 2004, 
ApJ, 610, 876

\bibitem[]{893}
Kawabata, K.S., et al. 2003, ApJ, 593, L19


\bibitem[]{897}
Kobayashi, C., et al. 2006, ApJ, submitted

\bibitem[]{900}
Kozma, C., et al. 2005, A\&A, 437, 983

\bibitem[]{903}
Kumagai, S.,Shigeyama, T., Nomoto, K., Itoh, M., 
Nishimura, J., \& Tsuruta, S. 1989, ApJ, 345, 412

\bibitem[]{}
Leonard, D.C., Filippenko, A.V., \& Ardila, D.R. 
2001, ApJ, 553, 861

\bibitem[]{907}
Lucy, L.B. 2005, A\&A, 429, 19 

\bibitem[]{910}
MacFadyen, A. I., \& Woosley, S. E. 1999, ApJ, 524, 262

\bibitem[]{913} 
MacFadyen, A.I. 2003, 
in 'From Twilight to Highlight: The Physics of Supernovae', 
eds. W. Hillebrandt \& B. Leibundgut (Berlin: Springer), 97


\bibitem[]{919}
Maeda, K., Nakamura, T., Nomoto, K.,
Mazzali, P.A., Patat, F., \& Hachisu, I. 2002, ApJ, 565, 405

\bibitem[]{923}
Maeda, K., Mazzali, P.A., Deng, J., Nomoto, K., Yoshii, Y., Tomita, H., 
\& Kobayashi, Y. 2003a, ApJ, 593, 931

\bibitem[]{927}
Maeda, K., \& Nomoto, K., 2003b, ApJ, 598, 1163 

\bibitem[]{930}
Maeda, K., Nomoto, K., Mazzali, P.A., \& Deng, J. 2006a, 
ApJ, 640, in press (astro-ph/0508373)

\bibitem[]{}
Maeda, K. 2006b, ApJ, 644, in press (astro-ph/0511480)

\bibitem[]{934}
Matheson, T., Filippenko, A.V., Leonard, D.C., \& Shields, J.C. 2001, 
AJ, 121, 1648 

\bibitem[]{938}
Matheson, T., et al. 2003, ApJ, 599, 394

\bibitem[]{941}
Mazzali, P.A., Iwamoto, K., \& Nomoto, K. 2000, ApJ, 545, 407 

\bibitem[]{944}
Mazzali, P.A., Nomoto, K., Patat, F., 
\& Maeda, K. 2001, ApJ, 559, 1047

\bibitem[]{948}
Mazzali, P.A., et al. 2002, ApJ, 572, L61 

\bibitem[]{951}
Mazzali, P.A., et al. 2003, ApJ, 599, L95

\bibitem[]{954}
Mazzali, P.A., et al. 2005, Science, 308, 1284 

\bibitem[]{957}
McCray, R., Shull, J.M., \& Sutherland, P. 1987, ApJ, 317, L73

\bibitem[]{960}
McKenzie, E.H., \& Schaefer, B.E. 1999, PASP, 111, 964

\bibitem[]{963}
Milne, P.A., et al. 2004, ApJ, 613, 1101

\bibitem[]{966}
Nakamura, T., Mazzali, P. A., Nomoto, K., 
\& Iwamoto, K. 2001, ApJ, 550, 991

\bibitem[]{970} 
Nagataki, S. 2000, ApJS, 127, 141

\bibitem[]{973}
Nomoto, K., Thilelmann, F.-K., \& Yokoi, K. 1984, 
ApJ, 286, 644

\bibitem[]{977}
Nomoto, K., \& Hashimoto, M. 1988, Phys. Rep., 256, 173

\bibitem[]{980}
Nomoto, K., et al. 1994, Nature, 371, 227

\bibitem[]{983} 
Nomoto, K., Maeda, K, Mazzali, P. A., Umeda, H., Deng, J., \& Iwamoto, K. 2004, 
in Stellar Collapse, ed. C. L. Fryer (Kluwer: Dordrecht), 277 
(astro-ph/0308136)


\bibitem[]{989}
Patat, F., et al. 2001, ApJ, 555, 900

\bibitem[]{992}
Podsiadlowski, Ph., Nomoto, K., 
Maeda, K., Nakamura, T., Mazzali, P.A., \& Schmidt, B. 
2002, ApJ, 567, 491

\bibitem[]{997}
Proga, D., MacFadyen, A.I., Armitage, P.J., \& Begelman, M.C. 2003, 
ApJ, 599, 5

\bibitem[]{1001} 
Pruet, J., Surman, R., \& McLaughlin, G.C. 
2004a, ApJ, 602, L101

\bibitem[]{1005}
Pruet, J., Thompson, T.A., \& Hoffman, R.D. 
2004, ApJ, 606, 1006

\bibitem[]{1009}
R\"opke, F.K., \& Hillebrandt, W. 2005, A\&A, 431, 635

\bibitem[]{1012}
Sawai, H., Kotake, K., \& Yamada, S. 2005, ApJ, in press (astro-ph/0505611)

\bibitem[]{1015}
Sekiguchi, Y., \& Shibata, M. 2005, Phys.Rev. D71, 084013

\bibitem[]{1018}
Shibazaki, N., \& Ebisuzaki, T. 1988, ApJ, 327, L9 

\bibitem[]{1021}
Sollerman, J., Kozma, C., Fransson, C., Leibundgut, B., Lundqvist, P., 
Ryde, F., \& Woudt, P. 2000, ApJ, 537, L127

\bibitem[]{1025}
Stanek, K. Z., et al. 2003, ApJ, 591, L17

\bibitem[]{1028}
Yamazaki, R., Yonetoku, D., \& Nakamura, T. 2003, ApJ, 594, L79

\bibitem[]{}
Wang, L., Howell, D.A., H\"oflich, P., \& Wheeler, J.C. 
2001, ApJ, 550, 1030 

\bibitem[]{}
Wang, L., et al. 2002, ApJ, 579, 671 

\bibitem[]{1031}
Wheeler, J.C., Yi, I., P., \& Wang, L. 2000, ApJ, 537, 810

\bibitem[]{1034}
Woosley, S.E., Pinto, P.A., Martin, P.G., 
\& Weaver, T.A. 1987, ApJ, 318, 664 

\bibitem[]{1038}
Woosley, S.E., Eastman, E.G., \& Schmidt, B.P. 1999, ApJ, 516, 788




\end{thebibliography}
\end{document}